  \pdfoutput=1
  \documentclass[graphicx,amssymb,amsmath,enumerate,11pt]{report}
  \usepackage{amssymb}
\usepackage{fancyhdr}
  \fancyheadoffset[]{0.1cm}

  \usepackage{epsfig}
  \usepackage{url}
  \newcommand\mchapter[2]{\chapter*{#1}
  \vskip -0.5cm \noindent {\it \LARGE #2}
  \addcontentsline{toc}{chapter}{#1\\{\normalsize\it #2}}}

\begin{document}      

 \rhead{\bfseries Neutrino mass from Cosmology}

 \mchapter{Neutrino mass from Cosmology}
 {Julien Lesgourgues\,$^a$ and Sergio Pastor\,$^b$}
 \label{ch-20:numasscosmo}

\vspace{0.5cm}

\begin{center}
$^a$ {\it CERN, Theory Division, CH-1211 Geneva 23, Switzerland \\
                Institut de Th{\'e}orie des Ph{\'e}nom{\`e}nes Physiques, EPFL, CH-1015 Lausanne, Switzerland\\
                LAPTH (CNRS-Universit{\'e} de Savoie), B.P. 110, F-74941 Annecy-le-Vieux Cedex, France} \\ [6pt]
$^b$ {\it Instituto de F\'{\i}sica Corpuscular  (CSIC-Universitat de Val\`{e}ncia), \\
Apdo.\ correos 22085, 46071 Valencia, Spain}
\end{center}

\vspace{1cm}

\begin{center}
{\bf Abstract}
\end{center}
Neutrinos can play an important role in the evolution of the Universe,
modifying some of the cosmological observables. In this contribution
we summarize the main aspects of cosmological relic neutrinos and we
describe how the precision of present cosmological data can be used to
learn about neutrino properties, in particular their mass, providing
complementary information to beta decay and neutrinoless double-beta
decay experiments. We show how the analysis of current cosmological
observations, such as the anisotropies of the cosmic microwave
background or the distribution of large-scale structure, provides an
upper bound on the sum of neutrino masses of order 1 eV or less, with 
very good perspectives
from future cosmological measurements which are expected to be
sensitive to neutrino masses well into the sub-eV range.

\vskip 1cm

\section{Introduction}
\label{20-sec:intro}

The subject of this contribution is the role of neutrinos in
Cosmology, one of the best examples of the very close ties that have
developed between nuclear physics, particle physics, astrophysics and
cosmology. Here we focus on the most interesting aspects related to 
the case of massive (and light) relic neutrinos, but many
others that were left out can be found in \cite{20-Dolgov:2002wy,20-NuCosmo}.

We begin with a description of the properties and evolution of the
background of relic neutrinos that fills the Universe. Then we review
the possible effects of neutrino oscillations on Cosmology.  
The topic of neutrinos and Big Bang Nucleosynthesis is reviewed
in a different contribution to this Special Issue \cite{20-Steigman}.
The largest part of this contribution is devoted to the impact of massive
neutrinos on cosmological observables, that can be used to extract
bounds on neutrino masses from present data. Finally we discuss the
sensitivities on neutrino masses from future cosmological experiments.

Note that massive neutrinos could also play a role in the
generation of the baryon asymmetry of the Universe from a previously
created lepton asymmetry. In these leptogenesis scenarios, one can
also obtain quite restrictive bounds on light neutrino masses, which
are however strongly model-dependent. We do not discuss this subject
here, covered in other contribution to this Special Issue \cite{20-Nardi}.

For further details, the reader is referred to recent reviews on
neutrino cosmology such as 
\cite{20-Hannestad:2006zg,20-Hannestad:2010kz,20-Wong:2011ip} and in particular 
\cite{20-Lesgourgues:2006nd}.  A more general review on the connection between
particle physics and cosmology can be found in
\cite{20-Kamionkowski:1999qc}.

\section{The cosmic neutrino background}
\label{20-sec:theCNB}

The existence of a relic sea of neutrinos is a generic feature of the
standard hot big bang model, in number only slightly below that of
relic photons that constitute the cosmic microwave background
(CMB). This cosmic neutrino background (CNB) has not been detected
yet, but its presence is indirectly established by the accurate
agreement between the calculated and observed primordial abundances of
light elements, as well as from the analysis of the power spectrum of
CMB anisotropies and other cosmological observables.  In this section we will summarize the evolution and
main properties of the CNB.

\subsection{Relic neutrino production and decoupling}
\label{20-subsec:nudec}

Produced at large temperatures by frequent weak interactions, cosmic
neutrinos of any flavour ($\nu_{e,\mu,\tau}$) were kept in
equilibrium until these processes became ineffective in the course of
the expansion of the early Universe. While coupled to the rest of the
primeval plasma (relativistic particles such as electrons, positrons
and photons), neutrinos had a momentum spectrum with an equilibrium
Fermi-Dirac form with temperature $T$,
\begin{equation}
f_{\rm eq}(p,T)=\left
[\exp\left(\frac{p-\mu_\nu}{T}\right)+1\right]^{-1}\,,
\label{20-FD}
\end{equation}
which is just one example of the general case of particles in
equilibrium (fermions or bosons, relativistic or non-relativistic), as
shown e.g.\ in \cite{20-kt}. In the previous equation we have included a
neutrino chemical potential $\mu_\nu$ that would exist in the presence
of a neutrino-antineutrino asymmetry, but we will see later in section
\ref{20-sec:active-active} that even if it exists its contribution can not be
very relevant.

As the Universe cools, the weak interaction rate $\Gamma_\nu$ falls
below the expansion rate and one says that neutrinos decouple from the
rest of the plasma. An estimate of the decoupling temperature $T_{\rm
dec}$ can be found by equating the thermally averaged value of the
weak interaction rate
\begin{equation}
\Gamma_\nu=\langle\sigma_\nu\,n_\nu\rangle \,\, ,
\label{20-Gamma}
\end{equation}
where $\sigma_\nu \propto G_F^2$ is the cross section of the
electron-neutrino processes with $G_F$ the Fermi constant and
$n_\nu$ is the neutrino number density, with the expansion rate
given by the Hubble parameter $H$ 
\begin{equation}
H^2=\frac{8\pi\rho}{3M_P^2} \,\, .
\label{20-H_MeV}
\end{equation}
Here $\rho\propto T^4$ is the total energy density, dominated by
relativistic particles, and $M_P=1/G^{1/2}$ is the Planck mass.  If we
approximate the numerical factors to unity, with $\Gamma_\nu \approx
G_F^2T^5$ and $H \approx T^2/M_P$, we obtain the rough estimate
$T_{\rm dec}\approx 1$ MeV.  More accurate calculations give slightly
higher values of $T_{\rm dec}$ which are flavour dependent because
electron neutrinos and antineutrinos are in closer contact with
electrons and positrons, as shown e.g.\ in \cite{20-Dolgov:2002wy}.

Although neutrino decoupling is not described by a unique $T_{\rm
dec}$, it can be approximated as an instantaneous process.  The
standard picture of {\it instantaneous neutrino decoupling} is very
simple (see e.g.\ \cite{20-kt} or \cite{20-dodelson}) and
reasonably accurate.  In this approximation, the spectrum in eq.\
(\ref{20-FD}) is preserved after decoupling, because both neutrino momenta
and temperature redshift identically with the expansion of the
Universe. In other words, the number density of non-interacting
neutrinos remains constant in a comoving volume since the decoupling
epoch.  We will see later that active neutrinos cannot possess masses
much larger than $1$ eV, so they were ultra-relativistic at
decoupling. This is the reason why the momentum distribution in eq.\
(\ref{20-FD}) does not depend on the neutrino masses, even after
decoupling, i.e.\ there is no neutrino energy in the exponential of
$f_{\rm eq}(p)$.

When calculating quantities related to relic neutrinos, one must
consider the various possible degrees of freedom per flavour. If
neutrinos are massless or Majorana particles, there are two degrees of
freedom for each flavour, one for neutrinos (one negative helicity
state) and one for antineutrinos (one positive helicity state).
Instead, for Dirac neutrinos there are in principle twice more degrees
of freedom, corresponding to the two helicity states. However, the
extra degrees of freedom should be included in the computation only if
they are populated and brought into equilibrium before the time of
neutrino decoupling. In practice, the Dirac neutrinos with the
``wrong-helicity'' states do not interact with the plasma at
temperatures of the MeV order and have a vanishingly small density
with respect to the usual left-handed neutrinos (unless neutrinos have
masses close to the keV range, as explained in section 6.4 of
\cite{20-Dolgov:2002wy}, but such a large mass is excluded for
active neutrinos). Thus the relic density of active neutrinos does not
depend on their nature, either Dirac or Majorana particles.

\begin{figure}[t]
\includegraphics[width=.68\textwidth]{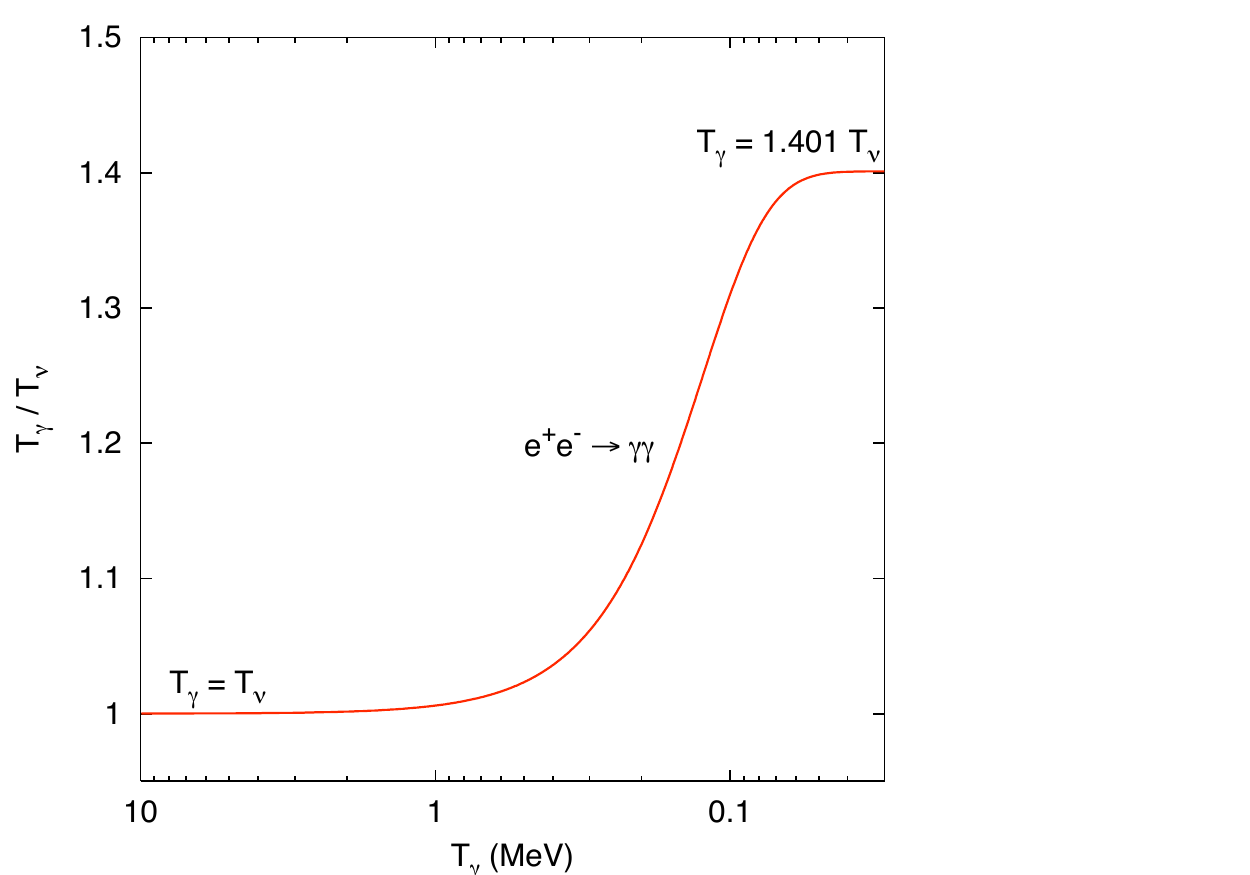}
\hspace{-3.2cm}
\includegraphics[width=.68\textwidth]{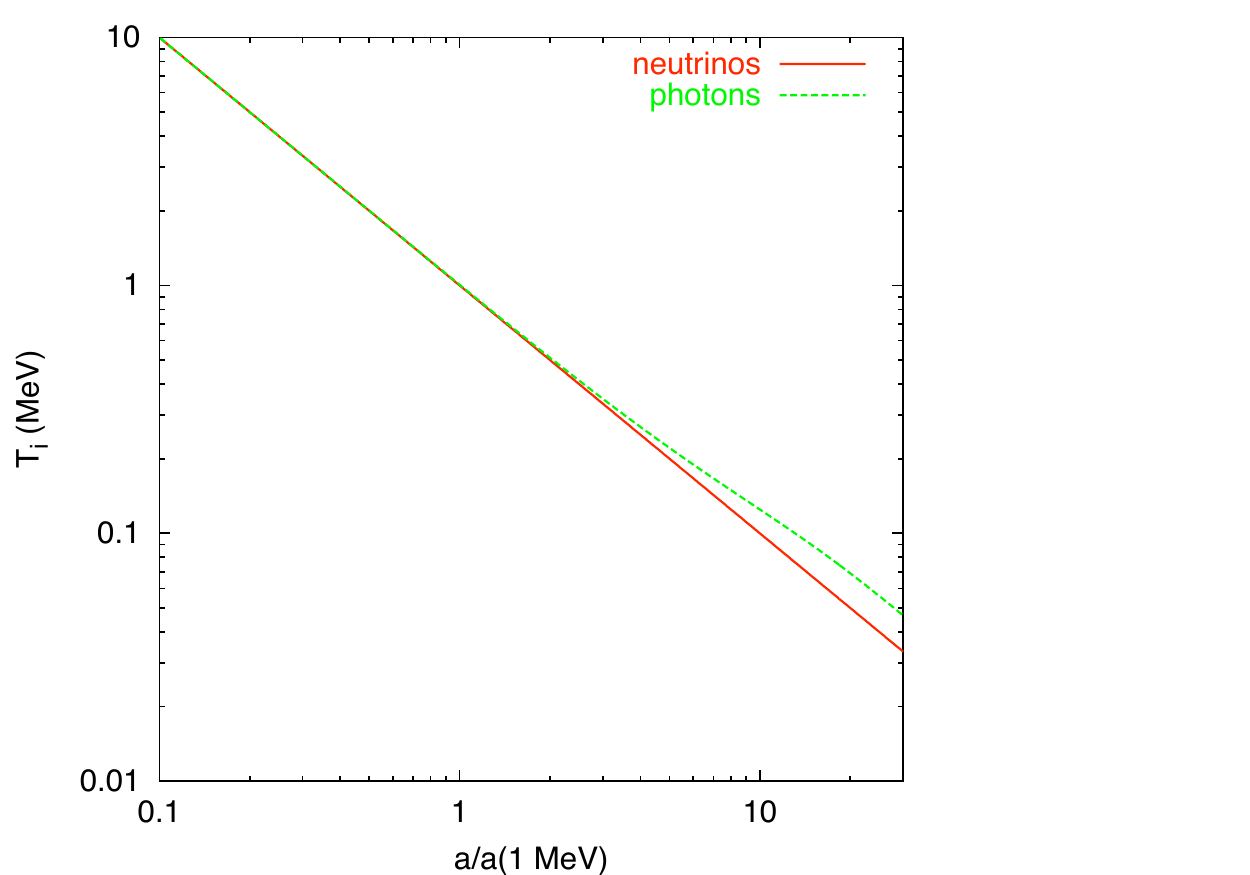}
\caption{\label{20-fig:tnu} Photon and neutrino temperatures during the
process of $e^{\pm}$ annihilations: evolution of their ratio (left)
and their decrease with the expansion of the Universe (right).
}
\end{figure}

Shortly after neutrino decoupling the temperature drops below the
electron mass, favouring $e^{\pm}$ annihilations that heat the
photons. If one assumes that this entropy transfer did not affect the
neutrinos because they were already completely decoupled, it is easy
to calculate the change in the photon temperature before any $e^{\pm}$
annihilation and after the electron-positron pairs disappear by
assuming entropy conservation of the electromagnetic plasma. The
result is
\begin{equation}
\frac{T^{\rm after}_\gamma}{T^{\rm before}_\gamma}=\left
(\frac{11}{4}\right)^{1/3}\simeq 1.40102\, ,
\label{20-Tgamma_e+e-}
\end{equation}
which is also the ratio between the temperatures of relic photons and
neutrinos $T_\gamma/T_\nu=(11/4)^{1/3}$. The evolution of this ratio
during the process of $e^{\pm}$ annihilations is shown in the left
panel of Fig.\ \ref{20-fig:tnu}, while one can see in the right panel how
in this epoch the photon temperature decreases with the expansion less
than the inverse of the scale factor $a$. Instead the temperature of
the decoupled neutrinos always falls as $1/a$.

It turns out that the standard picture of neutrino decoupling
described above is slightly modified: the processes of neutrino
decoupling and $e^{\pm}$ annihilations are sufficiently close in time
so that some relic interactions between $e^{\pm}$ and neutrinos exist.
These relic processes are more efficient for larger neutrino energies,
leading to non-thermal distortions in the neutrino spectra at the percent level and a slightly smaller increase of the comoving photon
temperature, as noted in a series of works listed in \cite{20-Dolgov:2002wy,20-NuCosmo}.
A proper calculation of the process
of non-instantaneous neutrino decoupling demands solving the
momentum-dependent Boltzmann equations for the neutrino spectra, a set
of integro-differential kinetic equations that are difficult to solve
numerically. This problem was considered in \cite{20-Mangano:2005cc}
including the effect of flavour neutrino oscillations on
the neutrino decoupling process. One finds an increase in the neutrino
energy densities with respect to the instantaneous decoupling
approximation ($0.73\%$ and $0.52\%$ for $\nu_e$'s and
$\nu_{\mu,\tau}$'s, respectively) and a value of the comoving photon
temperature after $e^\pm$ annihilations which is a factor $1.3978$
larger, instead of $1.40102$. These changes modify the contribution of
relativistic relic neutrinos to the total energy density which is
taken into account using $N_{\rm eff} \simeq 3.046$, as defined later
in eq.\ (\ref{20-neff}). In practice, the distortions calculated in
\cite{20-Mangano:2005cc} only have small consequences on the
evolution of cosmological perturbations, and for many purposes they
can be safely neglected.

Any quantity related to relic neutrinos can be calculated after
decoupling with the spectrum in eq.\ (\ref{20-FD}) and $T_\nu$. For
instance, the number density per flavour is fixed by the temperature,
\begin{equation}
n_{\nu} = \frac{3}{11}\;n_\gamma =
\frac{6\zeta(3)}{11\pi^2}\;T_\gamma^3~,
\label{20-nunumber}
\end{equation}
which leads to a present value of $113$ neutrinos and antineutrinos of
each flavour per cm$^{3}$. Instead, the energy density for massive
neutrinos should in principle be calculated numerically, with two
well-defined analytical limits,
\begin{eqnarray}
\rho_\nu (m_\nu \ll T_\nu) & = & 
\frac{7\pi^2}{120}
\left(\frac{4}{11}\right)^{4/3}\;T_\gamma^4~,
\label{20-rhomassless}\\
\rho_\nu (m_\nu \gg T_\nu) & = & m_\nu n_\nu~.
\label{20-nurho}
\end{eqnarray}

\subsection{Background evolution}
\label{20-subsec:background}

Let us discuss the evolution of the CNB after decoupling in the
expanding Universe, which is described by the
Friedmann-Robertson-Walker metric \cite{20-dodelson}
\begin{equation}
ds^2 = dt^2 - a(t)^2\,\delta_{ij} dx^i dx^j ~,
\label{20-Friedmann_metric}
\end{equation}
where we assumed negligible spatial curvature. Here $a(t)$ is the
scale factor usually normalized to unity now ($a(t_0)=1$) and related
to the redshift $z$ as $a=1/(1+z)$.  General relativity tells us the
relation between the metric and the matter and energy in the Universe
via the Einstein equations, whose time-time component is the Friedmann
equation
\begin{equation}
\left(\frac{\dot{a}}{a}\right )^2=H^2 = \frac{8 \pi G}{3} 
\rho= H_0^2\frac{\rho}{\rho_{\rm c}^0}~,
\end{equation}
that gives the Hubble rate in terms of the total energy density
$\rho$.  At any time, the critical density $\rho_{\rm c}$ is defined
as $\rho_{\rm c}=3H^2/8\pi G$, and the current value $H_0$ of the
Hubble parameter gives the critical density today
\begin{equation}
\label{20-critical density}
\rho_{\rm c}^0 = 1.8788 \times 10^{-29}\,h^2 ~\mathrm{g~cm}^{-3}~.
\end{equation}
where $h\equiv H_0/(100~{\rm km\, s}^{-1} \,{\rm Mpc}^{-1})$.
The different contributions to the total energy density are
\begin{equation}
\rho=
\rho_{\gamma} + \rho_{\rm cdm} + \rho_{\rm b} 
+  \rho_{\nu} + \rho_{\Lambda}~,
\label{20-rhotot}
\end{equation}
and the evolution of each component is given by the energy
conservation law in an expanding Universe $\dot{\rho}=-3H(\rho+p)$,
where $p$ is the pressure. Thus the homogeneous density of photons
$\rho_{\gamma}$ scales like $a^{-4}$, that of non-relativistic matter
($\rho_{\rm cdm}$ for cold dark matter and $\rho_{\rm b}$ for baryons)
like $a^{-3}$, and the cosmological constant density $\rho_{\Lambda}$
is of course time-independent. Instead, the energy density of 
neutrinos contributes to the radiation density at early times but they
behave as matter after the non-relativistic transition. 

\begin{figure}[t]
\begin{center}
\includegraphics[width=.5\textwidth]{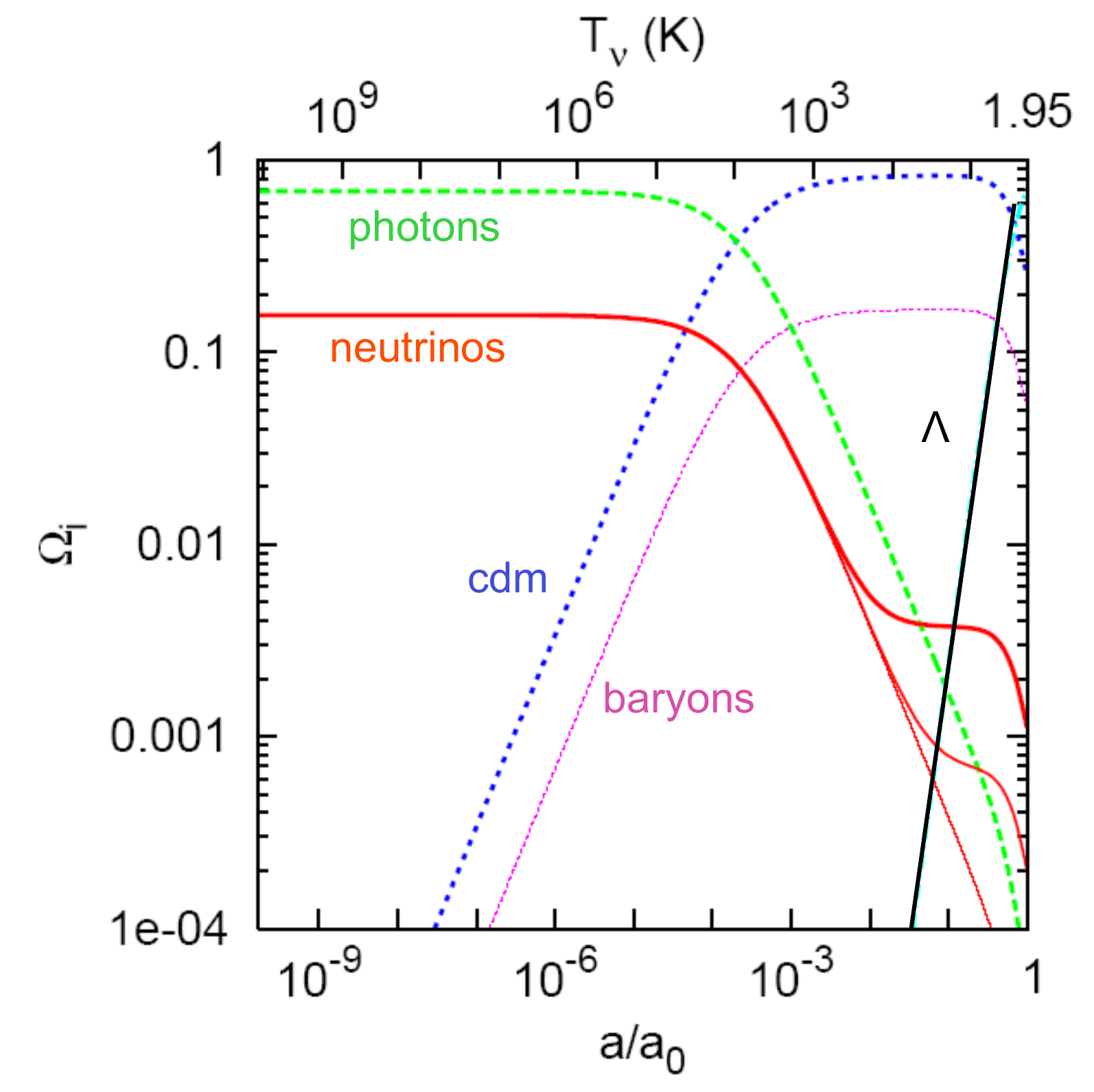}
\caption{\label{20-fig:rho_i} Evolution of the background energy
densities in terms of the fractions $\Omega_i$,
from the
time when $T_{\nu}=1$ MeV until now, for each component of a flat
Universe with $h=0.7$ and current density fractions
$\Omega_{\Lambda}=0.70$, $\Omega_{\rm b}=0.05$
and $\Omega_{\rm cdm}=1-\Omega_{\Lambda}-\Omega_{\rm b}
-\Omega_{\nu}$. The three neutrino masses are $m_1=0$, $m_2 = 0.009$
eV and $m_3 = 0.05$ eV.}
\end{center}
\end{figure}

The evolution of all densities is depicted in 
Fig.\ \ref{20-fig:rho_i}, starting at MeV temperatures until now. The
various density fractions $\Omega_{i}\equiv\rho
_{i}/\rho_{\rm c}$ are shown in this figure, where it is easy to
see which of the Universe components is dominant, fixing its expansion
rate: first radiation in the form of photons and neutrinos (Radiation
Domination or RD), then matter which can be CDM, baryons and massive
neutrinos at late times (Matter Domination or MD) and finally the
cosmological constant density takes over at low redshift (typically $z
< 0.5$).

Massive neutrinos are the only particles that present the transition
from radiation to matter, when their density is clearly enhanced
(upper solid lines in Fig.\ \ref{20-fig:rho_i}). Obviously the
contribution of massive neutrinos to the energy density in the
non-relativistic limit is a function of the mass (or the sum of all
masses for which $m_i \gg T_\nu$), and the present value $\Omega_\nu$
could be of order unity  for eV masses (see 
section \ref{20-sec:nuDM}).

Shortly after neutrino decoupling, the CNB plays an interesting role in
Big Bang Nucleosynthesis (BBN), the period of the Universe when the primordial
abundances of light elements are created. This subject is described in 
the contribution \cite{20-Steigman} (for a recent review, see
\cite{20-Iocco:2008va}). Here we just summarize the 
two main effects of relic neutrinos at BBN. The first one is
that they contribute to the relativistic energy density of the
Universe (if $m_\nu \ll T_\nu$), thus fixing the expansion rate. This
is why BBN gave the first allowed range of the number of neutrino
species before accelerators (see the next section).  On the other
hand, BBN is the last cosmological epoch sensitive to neutrino
flavour, because electron neutrinos and antineutrinos play a direct role
in the weak processes. We will see some examples
of BBN bounds on neutrinos (effective number or oscillations) in the
following sections.

\section{Extra radiation and the effective number of neutrinos}
\label{20-subsec:neff}

Together with photons, in the standard case neutrinos fix the
expansion rate during the cosmological era when the Universe is
dominated by radiation. Their contribution to the total radiation
content can be parametrized in terms of the effective number of
neutrinos $N_{\rm eff}$, through the relation
\begin{equation}
\rho_{\rm r} = \rho_\gamma + \rho_\nu = 
\left[ 1 + \frac{7}{8} \left( \frac{4}{11}
\right)^{4/3} \, N_{\rm eff} \right] \, \rho_\gamma \,\,,
\label{20-neff}
\end{equation}
where we have normalized $\rho_{\rm r}$ to the photon energy density because its
value today is known from the measurement of the CMB temperature. This
equation is valid when neutrino decoupling is complete and holds as
long as all neutrinos are relativistic. 

We know that the number of light neutrinos sensitive to weak
interactions (flavour or active neutrinos) equals three from the
analysis of the invisible $Z$-boson width at LEP, 
$N_\nu=2.9840 \pm
0.0082$ \cite{20-Yao:2006px}, and we saw in a previous section from the
analysis of neutrino decoupling that they contribute as $N_{\rm
eff}\simeq 3.046$. Any departure of $N_{\rm eff}$ from this last value
would be due to non-standard neutrino features or to the contribution
of other relativistic relics. For instance, the energy density of a
hypothetical scalar particle $\phi$ in equilibrium with the same
temperature as neutrinos would be $\rho_\phi=(\pi/30)\,T_\nu^4$,
leading to a departure of $N_{\rm eff}$ from the standard value of
$4/7$.  A detailed discussion of cosmological scenarios where $N_{\rm
eff}$ is not fixed to three can be found in \cite{20-Dolgov:2002wy,20-NuCosmo,20-Sarkar:1995dd}. 

The expansion rate during BBN
fixes the produced abundances of light elements, and in particular
that of $^4$He. Thus the value of $N_{\rm eff}$ can be constrained at
the BBN epoch from the comparison of theoretical predictions and
experimental data on the primordial abundances of light elements \cite{20-Steigman,20-Iocco:2008va}.  In
addition, a value of $N_{\rm eff}$ different from the standard one
would modify the transition epoch from a radiation-dominated to a
matter-dominated Universe, which has some consequences on some
cosmological observables such as the power spectrum of CMB
anisotropies, leading to independent bounds on the radiation content.
These are two complementary ways of constraining $N_{\rm eff}$ at very
different epochs. 

A recent analysis of the BBN constraints on $N_{\rm eff}$ can be found in
\cite{20-Mangano:2011ar} (see the references
therein for a list of recent works). The authors have
discussed a new and more conservative approach to derive BBN constraints on 
$N_{\rm eff}$, motivated by growing concerns on the reliability of 
astrophysical determinations of primordial $^4$He. 
According to \cite{20-Mangano:2011ar}, the extra radiation 
at the BBN epoch is limited to $\Delta N_{\rm eff}\leq1$ at 95\% C.L.
On the other hand, recent analyses of late cosmological observables
seem to favor $N_{\rm eff}>3$, with best-fit values of
order $4.3-4.4$, although with large errorbars as shown for instance in \cite{20-Komatsu:2010fb}.
The considered data include 
CMB temperature anisotropies and polarization, combined with other data such as 
measurements of the present value of the Hubble parameter (H$_0$), 
the power spectrum of Luminous Red Galaxies (LRG) or
distance measurements from the baryon acoustic oscillations (BAO) in the distribution of
galaxies. The allowed regions at 95\% C.L.\ from \cite{20-Komatsu:2010fb}
are  $2.8<N_{\rm eff}<5.9$ (WMAP+BAO+H$_0$) and
$2.7<N_{\rm eff}<6.2$ (WMAP+LRG+H$_0$). Other recent 
analyses of $N_{\rm eff}$ bounds
from cosmological data can be found in 
\cite{20-Dunkley:2010ge,20-Keisler:2011aw,20-Hamann:2011hu}.
These ranges are in reasonable agreement with the standard prediction
of $N_{\rm eff}\simeq 3.046$, although they show a marginal preference 
for extra relativistic degrees of freedom whose robustness is still unclear.
Moreover, they show that there exists an
allowed region of $N_{\rm eff}$ values that is common at early (BBN)
and more recent epochs, although with large errorbars. The upcoming CMB 
measurements by the {\sc Planck} satellite will soon pin
down the radiation content of the Universe, clarifying whether one 
really needs new physics leading to
relativistic degrees of freedom beyond the
contribution of flavour neutrinos. 

\section{Neutrino oscillations in the Early Universe}
\label{20-sec:osci}

Nowadays there exist compelling evidences for flavour neutrino
oscillations from a variety of experimental data on solar,
atmospheric, reactor and accelerator neutrinos. These are very
important results, because the existence of flavour change implies
that neutrinos mix and have non-zero masses, which in turn requires
particle physics beyond the Standard Model. Thus it is interesting to
check whether neutrino oscillations can modify any of the cosmological
observables.  More on neutrino oscillations and their implications can
be found in other contributions to this Special Issue or any of the existing reviews such as
\cite{20-Fogli:2005cq,20-GonzalezGarcia:2007ib,20-Schwetz:2011qt}, to which we
refer the reader for more details.

It turns out that in the standard cosmological picture all flavour
neutrinos were produced with the same energy spectrum, as we saw in
section \ref{20-subsec:nudec}, so we do not expect any effect from the
oscillations among these three states. This is true up to the small
spectral distortion caused by the heating of neutrinos from $e^+e^-$
annihilations \cite{20-Mangano:2005cc}, as  described before. In this
section we will briefly consider two cases where neutrino oscillations
could have cosmological consequences: flavour oscillations with
non-zero relic neutrino asymmetries and active-sterile neutrino
oscillations.  

\subsection{Active-active neutrino oscillations: relic neutrino asymmetries}
\label{20-sec:active-active}

A non-zero relic neutrino asymmetry exists when the number densities
of neutrinos and antineutrinos of a given flavour are different, and can
be parameterized by the ratio $\eta_{\nu_\alpha} = 
(n_{\nu_\alpha}-n_{\overline{\nu}_\alpha})/n_\gamma$.  Such
a putative asymmetry, that could have been produced by some mechanism 
well before the thermal decoupling epoch, 
is expected to be of the
same order of the cosmological  baryon number $\eta_B$, i.e. 
a few times $10^{-10}$, from the equilibration of lepton and baryon numbers 
by sphalerons in the very early Universe. In such a case,
cosmological neutrino asymmetries would be too small to have any observable 
consequence. However, values fo these parameters which are 
orders of magnitude larger than $\eta_B$
are not excluded by observations. Actually, large $\eta_{\nu_\alpha}$ 
are predicted in theoretical models where the generation of lepton asymmetry took place
after the electroweak phase transition or the electroweak washing out of preexisting
asymmetries is not effective. 

Neutrino asymmetries can be
quantified assuming that a given flavour is characterized by a
Fermi-Dirac distribution as in eq.\ (\ref{20-FD}) with a non-zero
chemical potential $\mu_\nu$ or equivalently with the dimensionless
degeneracy parameter $\xi_\nu \equiv \mu_\nu/T$ (for antineutrinos,
$\xi_{\bar{\nu}}=-\xi_\nu$). In such a case, sometimes one says that
the relic neutrinos are degenerate (but not in the sense of equal
masses).
Degenerate electron neutrinos have a direct effect on BBN:  any change in the $\nu_e/\bar{\nu}_e$
spectra modifies the primordial neutron-to-proton ratio, which in this
case is $n/p\propto \exp(-\xi_{\nu_e})$.  Therefore, a positive
$\xi_{\nu_e}$ decreases the primordial $^4$He mass fraction, while a
negative $\xi_{\nu_e}$ increases it \cite{20-Kang:1991xa}, leading to an allowed range
\begin{equation}
-0.01<\xi_{\nu_e}<0.07~,
\label{20-stronglimit}
\end{equation}
compatible with $\xi_{\nu_e}=0$ and very restrictive for negative
values.
In addition a non-zero relic neutrino asymmetry always enhances the
contribution of the CNB to the relativistic energy density, because for
any $\xi_\nu$ one has a departure from the standard value of the
effective number of neutrinos $N_{\rm eff}$ given, if the neutrino
spectra follow an equilibrium form, by
\begin{equation}
\Delta N_{\rm eff} =
\frac{15}{7}\left [2\left(\frac{\xi_\nu}{\pi}\right)^2 
+\left(\frac{\xi_\nu}{\pi}\right)^4\right]
\label{20-neff_xi}
\end{equation}
We have seen that this increased radiation modifies the outcome of BBN
and that bounds on $N_{\rm eff}$ can be obtained. In addition, another
consequence of the extra radiation density is that it postpones the
epoch of matter-radiation equality, producing observable effects on
the spectrum of CMB anisotropies and the distribution of cosmic
large-scale structures (LSS). Both independent bounds on the radiation
content can be translated into flavour-independent limits on $\xi_\nu$.

Altogether these cosmological limits on the neutrino chemical
potentials or relic neutrino asymmetries are not very restrictive,
because at least for BBN their effect in the $\nu_\mu$ or $\nu_\tau$
sector can be compensated by a positive $\xi_{\nu_e}$. For example, an
analysis of the combined effect of a non-zero neutrino asymmetry on
BBN and CMB/LSS yields the allowed regions \cite{20-Hansen:2001hi}
\begin{equation}
-0.01 < \xi_{\nu_e} < 0.22, \qquad
|\xi_{\nu_{\mu,\tau}}| < 2.6,
\label{20-nooscbounds}
\end{equation}
in agreement with similar but more updated bounds as cited in
\cite{20-Lesgourgues:2006nd}.  These limits allow for a very significant
radiation contribution of degenerate neutrinos, leading many authors
to discuss the implications of a large neutrino asymmetry in different
physical situations (see e.g.\ \cite{20-Dolgov:2002wy}).

It is obvious that the limits in eq.\ (\ref{20-nooscbounds}) would be
modified if neutrino flavour oscillations were effective before BBN,
changing the initial distribution of flavour asymmetries. Actually, it was shown in
\cite{20-Dolgov:2002ab,20-Wong:2002fa,20-Abazajian:2002qx} that this is the case for the neutrino
mixing parameters in the region favoured by present data. This result
is obtained only after the proper inclusion of the refractive terms
produced by the background neutrinos, which synchronize the
oscillations of neutrinos with different momenta (which would evolve
independently without them).  If flavour equilibrium is
reached before BBN and the momentum distributions of neutrinos keep
a Fermi-Dirac form as in eq.\ (\ref{20-FD}), the restrictive 
limits on $\xi_{\nu_e}$ in eq.\
(\ref{20-stronglimit}) apply to all flavours. The bounds on the
common value of the neutrino degeneracy parameter $\xi_\nu\equiv
\mu_\nu/T$ would be $-0.05<\xi_\nu<0.07$ at $2\sigma$
\cite{20-Serpico:2005bc}. 

More recent analyses
\cite{20-Pastor:2008ti,20-Mangano:2010ei} have shown that
this conclusion does not always hold.
For the present measured values of neutrino mixing parameters, the degree of flavour
equilibration depends on the value of the mixing angle $\theta_{13}$, which fixes the onset of flavour 
oscillations involving $\nu_e$'s. This in turn
determines whether neutrinos interact enough with electrons and positrons to 
transfer the excess of energy density due to the initial $\eta_{\nu_\alpha}$
to the electromagnetic plasma. 

\begin{figure}
\begin{center}
  \includegraphics[width=0.5\textwidth,angle=0]{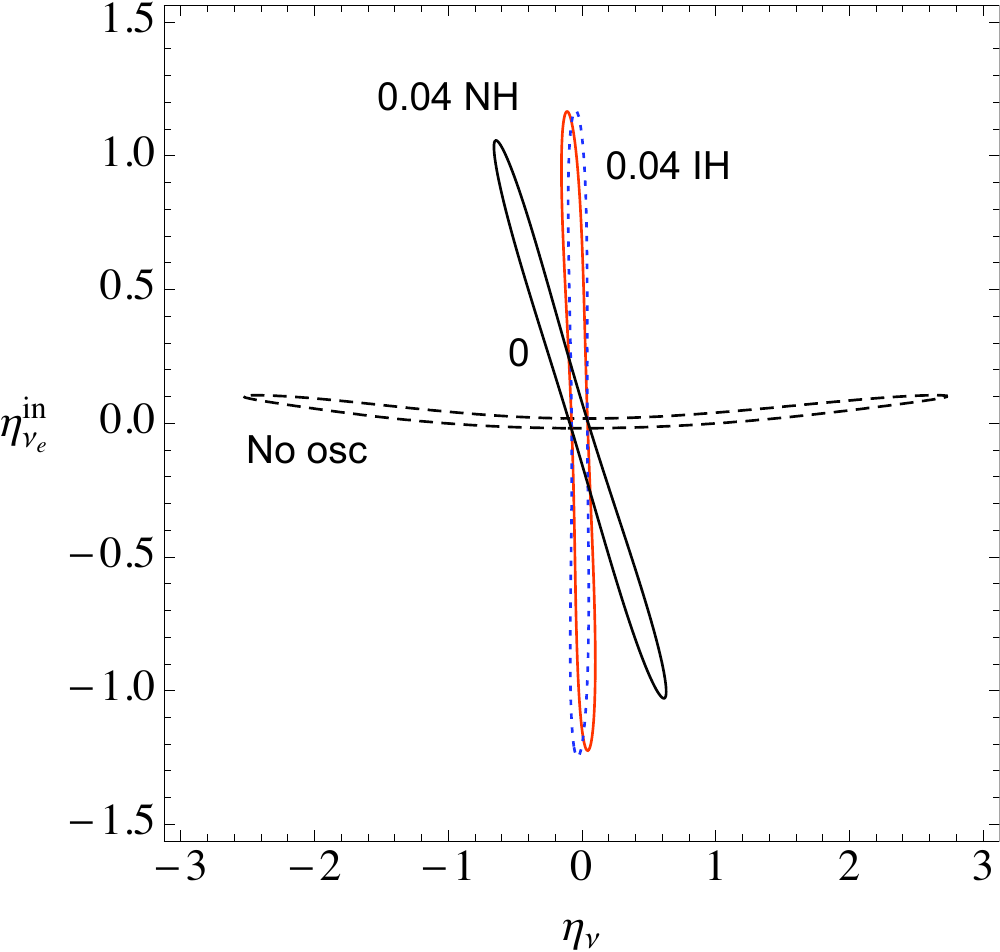}
   \end{center}
\caption{\label{20-bounds_eta} 
BBN contours at 95\% C.L.\ 
in the $\eta_\nu-\eta_{\nu_e}^{\rm in}$ plane for several values of
$\sin^2\theta_{13}$: 0 (solid line), 0.04 and normal mass hierarchy (NH) (almost vertical solid line), 0.04
and inverted mass hierarchy (IH) (dotted line)  \cite{20-Mangano:2011ip}.}
\end{figure}

In \cite{20-Mangano:2010ei}
the BBN bounds on the total neutrino asymmetry (cosmological lepton number) were
found for a range of initial flavour neutrino asymmetries.
An example of this analysis is shown in 
Fig.\ \ref{20-bounds_eta}, taken from \cite{20-Mangano:2011ip}. 
{}From this plot one can easily see the effect of flavour oscillations on
the BBN constraints on the total neutrino asymmetry. With no neutrino 
mixing the value of $\eta_{\nu_e}$ is severely constrained by 
$^4$He data, while
the asymmetry for other neutrino flavours could be much larger.
On the other hand, flavour oscillations implies that an initially  large
$\eta_{\nu_e}^{\rm in}$  can be compensated by 
an asymmetry in the other flavours with opposite sign. 
The most restrictive BBN bound on $\eta_{\nu_e}$ applies then to the total asymmetry, an
effect that can be seen graphically in Fig.\ \ref{20-bounds_eta}  as a {\it rotation}
of the allowed region from a quasi-horizontal one for zero mixing to an almost vertical
region for $\sin^2\theta_{13}=0.04$, in particular for the inverted mass hierarchy.
In all cases depicted in Fig.\  \ref{20-bounds_eta} the allowed values of the asymmetries are  fixed
by both deuterium and $^4$He, the latter imposing that the value of $\eta_{\nu_e}$ at BBN must be 
very close to zero. Finally, in \cite{20-Castorina:2012md} it has been shown that $^4$He data still
fix the bounds on $\eta_{\nu}$ when CMB results and other cosmological parameters such
as neutrino masses are included in the analysis, and only for future CMB data the bounds on the asymmetry
could be improved. At the same time the contribution of neutrino asymmetries to an
enhancement of radiation is limited to $N_{\rm eff}\lesssim 3.1-3.2$ for values of $\theta_{13}$
in the region allowed by oscillation data. 

\subsection{Active-sterile neutrino oscillations}
\label{20-sec:active-sterile}

In addition to the flavour or active neutrinos (three species as we saw
from accelerator data), there could also exist extra massive neutrino
states that are sterile, i.e.\ singlets of the Standard Model gauge
group and thus insensitive to weak interactions. 
Most of the current data on neutrino oscillations can be perfectly
explained with only the three active species, but 
there exist a few experimental results, sometimes called anomalies, that cannot be explained in this framework.
If neutrino oscillations are responsible for all the experimental data, a solution might require additional, sterile, neutrino species.
These kind of particles are predicted by many theoretical models beyond the SM, being
neutral leptons insensitive to weak 
interactions whose only interaction is gravitational. Their masses are 
usually heavy, while lighter sterile neutrinos are 
rarer but possible. Here we will briefly summarize these anomalies observed in neutrino experiments, whereas
an updated review of this subject is given in \cite{20-Abazajian:2012ys}.

The long-standing evidence (more than $3\sigma$) for $\bar{\nu}_\mu\to\bar{\nu}_e$ oscillations
comes from the LSND experiment. Its results pointed out a $\Delta m^2\sim 1$ eV$^2$, much larger than those 
required from a three-neutrino analysis, as shown later in eq.\ (\ref{20-oscpardef}). At the same time, the KARMEN experiment, 
very similar but not identical to LSND, provided no support for
such evidence, while a recent data release of the MiniBooNE experiment, designed to check the LSND results with larger distance $L$ and energy $E$ 
but similar ratio $L/E$, is still consistent with the LSND signal. In addition, an unexplained excess of electron-like
events is observed in MiniBooNE at low energies. 
The simultaneous interpretation of LSND (antineutrino) and MiniBooNE (neutrino and antineutrino) results
in terms of sterile neutrino oscillations requires CP-violation or some other exotic
scenarios, as reviewed in \cite{20-Abazajian:2012ys}.

A new anomaly supporting oscillations with sterile neutrinos appeared from a revaluation
of reactor antineutrino fluxes, which found a $3\%$ increase relative to previous flux calculations.
As a result, data from reactor neutrino experiments at very short distances can be interpreted as an apparent
deficit of $\bar{\nu}_e$. This is known as the reactor antineutrino anomaly and is again
compatible with sterile neutrinos having a $\Delta m^2 > 1$ eV$^2$. 
Finally, an independent experimental evidence for $\nu_e$ disappearance at very short baselines
exists from the Gallium radioactive source experiments GALLEX and SAGE.

The existence of all these experimental hints for sterile neutrinos and a mass scale at the eV is intriguing,
but so far a fully consistent picture has not emerged. Many analyses have been performed trying to
explain all experimental data with 1 or 2 additional sterile neutrinos, known as the $3+1$ or $3+2$ schemes, with
the corresponding additional mixing parameters. It seems that none of these schemes 
does describe well all data, as explained in detail in \cite{20-Abazajian:2012ys}, but for the topics of this review
the potential existence of oscillations into sterile neutrinos would lead to important cosmological 
consequences, such as extra radiation from fully or partly thermalized sterile neutrinos or a larger
Hot or Warm Dark Matter component. The required values of neutrino masses in 
these $3+1$ or $3+2$ scenarios are, as we will see in section \ref{20-sec:current}, in tension with
the current cosmological bounds. 

In any case, it is interesting to consider the main effects of additional sterile neutrino species in Cosmology.
Because sterile neutrinos are insensitive to weak interactions, they do not follow the behaviour
of active neutrinos and are not expected to be present in the early Universe at MeV temperatures. 
Even if they could interact through other kind of interactions, significantly weaker than the standard weak ones,
as predicted by extended particle physics models, they would have a thermal spectrum at very high 
temperatures but their density would have been strongly
diluted by many subsequent particle-antiparticle annihilations.
Therefore, barring the non-thermal production from additional physics beyond the SM,
the main way of obtaining a significant abundance of
sterile neutrinos is through their mixing with the active ones.

In principle, the cosmological evolution of the active-sterile neutrino system should be found solving the
corresponding Boltzmann kinetic equations for the density matrices.
There exist a vast literature on this subject, where different analyses considered several approximations. For a list of references,
see for instance \cite{20-Dolgov:2002wy,20-NuCosmo}. Although in general one should consider, at least, a $4\times 4$ mixing of three 
active neutrinos and one sterile species
(with 4 masses, 6 mixing angles and 3 CP-violating phases), let us first assume an admixture of one sterile state to electron neutrinos.
In the early Universe one expects
that neutrino oscillations could be effective when the vacuum oscillation term becomes larger than the 
main matter potential term from charged leptons at a temperature $T_c$. 
The evolution of the active-sterile neutrino system depends on the sign of $\Delta m^2$. For negative values
there could be resonant oscillations (or resonant production, RP). Instead, for $\Delta m^2>0$ one has the so-called
non-resonant production of sterile neutrinos (NRP). 
Comparing the value of $T_c$ with the decoupling temperature
of active neutrinos $T_{\nu D}\sim 2$ MeV, one obtains that for values 
\begin{equation}
\Delta m^2\lesssim 1.3\times10^{-7}\,{\rm eV}^2
\label{20-dm2_dec}
\end{equation}
active-sterile oscillations are effective after neutrino decoupling. In such a case, the total comoving number density of
active and sterile neutrinos will be constant because active neutrinos are no longer interacting with the rest of the primeval plasma. 
Correspondingly, distortions in the momentum spectra of neutrinos are expected which directly affect 
the production of $^4$He at BBN when the active neutrinos are of the electron
flavour (see e.g.\ \cite{20-Kirilova:2006wh}, also for the case of
non-zero initial $\nu_s$ abundance).
Instead, for much larger values of $\Delta m^2$
oscillations are effective before neutrino decoupling, when weak interactions are frequent. 
In this case the actual growth of the sterile neutrino population depends on the interplay between oscillations and 
interactions, while the energy spectrum of the 
active ones will be kept in equilibrium. Their combined contribution to radiation can be as large as $N_{\rm eff}=2$, depending
on the specific value of the mixing angle and on the sign of $\Delta m^2$.
This ``thermalization'' of the sterile neutrinos is a well-known phenomenon
that is very difficult to avoid unless the cosmological scenario is
drastically modified. 

In the NRP case, it was shown in \cite{20-Barbieri:1989ti} that the production probability of sterile neutrinos is
\begin{equation}
\Gamma_s =\langle \sin^22\theta_m \sin^2(\omega_{\rm osc}t)\Gamma_a\rangle
\label{20-gamma_s}
\end{equation}
where $\theta_m$ is the mixing angle in matter and $\omega_{\rm osc}$ the frequency of oscillations in the medium.
Here $\Gamma_a$ is the production rate of active neutrinos in the plasma, 
and the averaging is made over the thermal background. If the oscillation frequency is very high one can
substitute  $\sin^2(\omega_{\rm osc}t)\simeq 1/2$, and if $\theta_m$ is small one obtains
\begin{equation}
\Gamma_s \approx \theta_m^2 \Gamma_a
\label{20-gamma_s2}
\end{equation}
for a small number density of $\nu_s$ and active neutrinos close to equilibrium. Therefore, a thermal or close to
thermal population of sterile neutrinos is expected provided that such a production rate of $\nu_s$ is larger than
the expansion rate of the Universe, a condition that holds unless either $\Delta m^2$ or $\theta$ is very small.

\begin{figure}[t]
\includegraphics[width=0.5\textwidth]{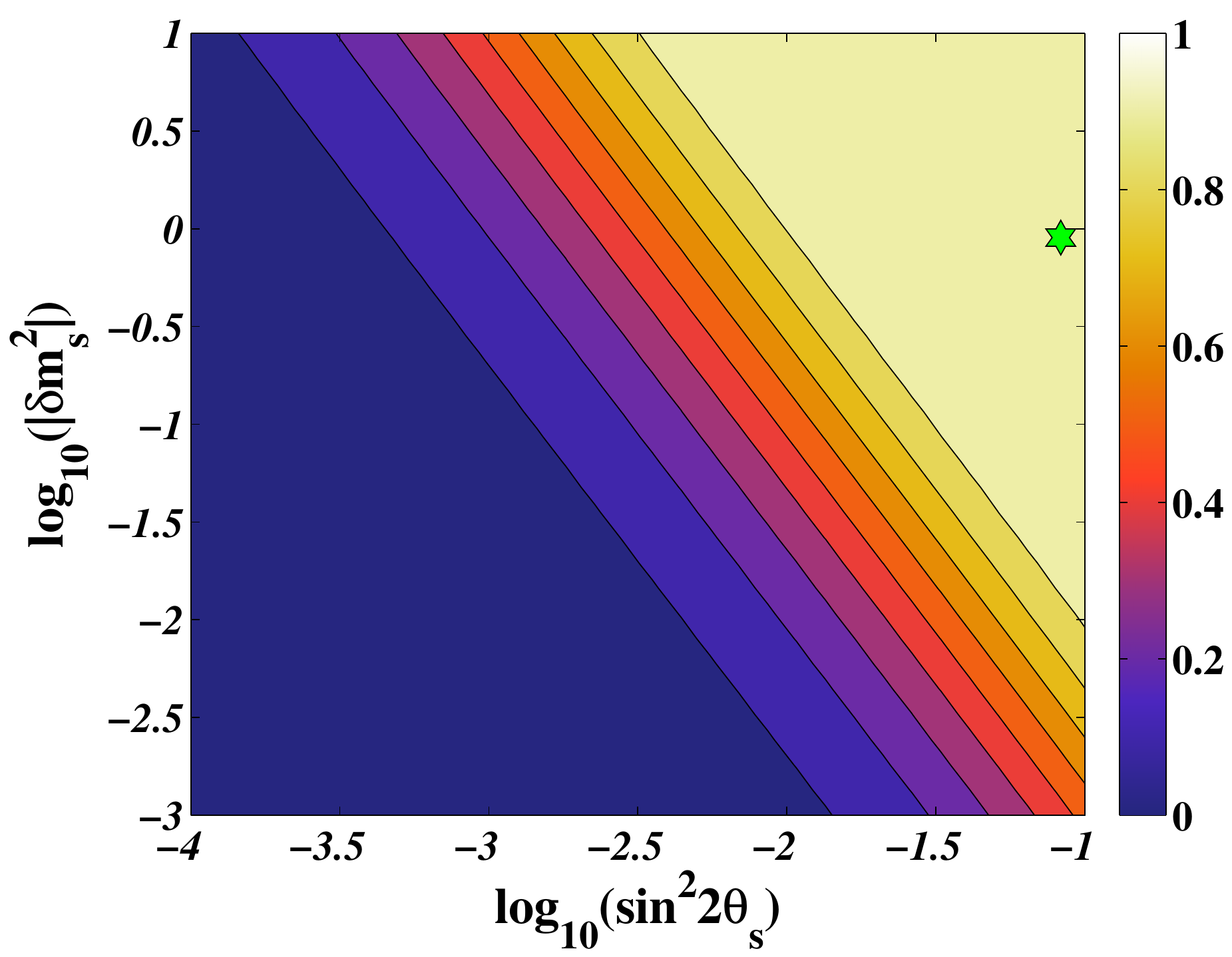}
\includegraphics[width=0.5\textwidth]{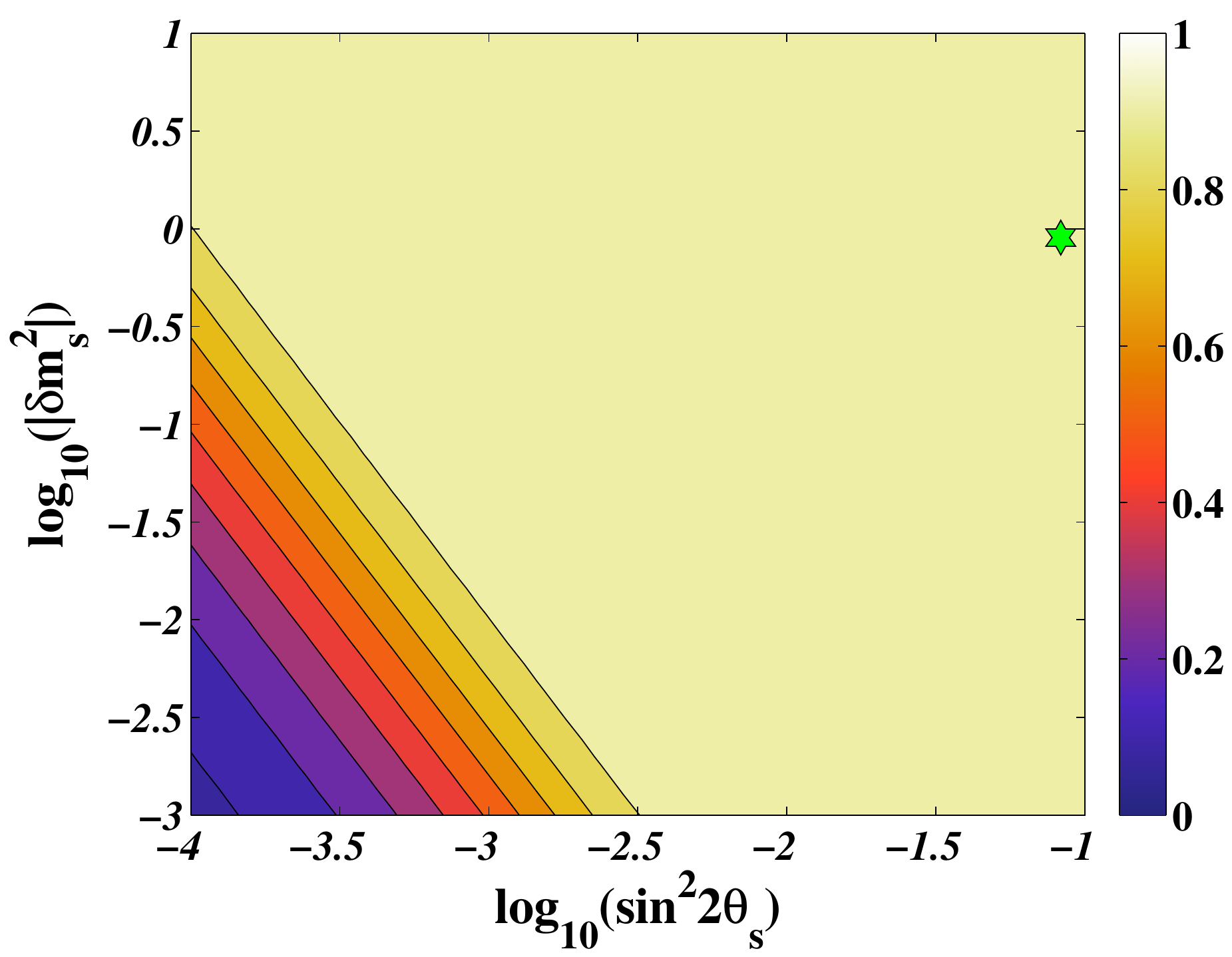}
\caption{\label{20-fig:therma_nus} Isocountours of the final value of $\Delta N_{\rm eff}$ 
in the $\sin^22\theta_s-\delta m_s^2$ plane for vanishing
lepton asymmetry and $\delta m_s^2>0$ (left panel) and
$\delta m_s^2<0$ (right pannel). The star denotes the best-fit mixing parameters as in the 3+1
global fit in \cite{20-Giunti:2011cp}: $(\delta m_s^2,\sin^22\theta_s)= (0.9~{\rm eV}^2, 0.089)$. Adapted from 
\cite{20-Hannestad:2012ky}. Courtesy of S.\ Hannestad, I.\ Tamborra, and T.\ Tram.}
\end{figure}

For RP of sterile neutrinos at temperatures below $T_c$ the resonance propagates from small to large values of 
the neutrino comoving momentum, covering the whole momentum distribution while the active neutrinos
are repopulated by interactions. The thermalization of $\nu_s$
is thus significantly enhanced, even for quite small values of the mixing angle.

In order to illustrate this discussion with an actual calculation of the active-sterile system with the kinetic equations
in the two-flavour approximation, among the many published analyses we
have chosen a recent one \cite{20-Hannestad:2012ky}. Their results for the final extra contribution of sterile neutrinos to radiation,
in the case of zero initial lepton asymmetry, are shown as isocontours of $\Delta N_{\rm eff}$
in Fig.\ \ref{20-fig:therma_nus} as a function of the mixing parameters $\delta m_s	^2\equiv \Delta m^2$ (in eV$^2$ units) 
and $\sin^22\theta_s\equiv \sin^22\theta$. In the NRP case (right panel) one can clearly see that the 
the same final $N_{\rm eff}$ corresponds to constant values of $\sin^42\theta\Delta m^2$. For RP
$\nu_s$'s are more efficiently brought into equilibrium, even for quite small values of the mixing angle.
All calculations described so far correspond to the two-neutrino limit of one active and one sterile states, but
a proper calculation should also include the unavoidable presence of mixing among the active neutrinos. A full
four-flavour calculation has not been performed, but a few analyses did solve simplified kinetic equations
taking into account active neutrino mixing, such as \cite{20-Dolgov:2003sg,20-Cirelli:2004cz,20-Mirizzi:2012we}.

We note that, for the active-sterile parameters needed to solve the oscillation anomalies described at the beginning of 
this section, the thermalization of sterile neutrinos is achieved, i.e.\ $N_{\rm eff}=4$. An example of 
the best-fit values of a particular calculation in the $3+1$ neutrino model is indicated in the plot. Therefore,
it seems that such an extra radiation is guaranteed in these situations unless oscillations are suppressed, as in the 
case of a non-zero initial lepton asymmetry $\eta_\nu$ much larger than $\eta_B$ \cite{20-Foot:1995bm,20-Kishimoto:2008ic}. 

\section{Massive neutrinos as Dark Matter}
\label{20-sec:nuDM}

Nowadays the existence of Dark Matter (DM), the dominant non-baryonic
component of the matter density in the Universe, is well established.
A priori, massive neutrinos are excellent DM candidates, in particular
because we are certain that they exist, in contrast with other
candidate particles. Together with CMB photons, relic neutrinos can be
found anywhere in the Universe with a number density given by the
present value of eq.\ (\ref{20-nunumber}) of $339$ neutrinos and
antineutrinos per cm$^{3}$, and their energy density in units of the
critical value of the energy density (see eq.\ (\ref{20-critical
density})) is
\begin{equation}
\Omega_{\nu} = \frac{\rho_\nu}{\rho^0_{\rm c}} =
\frac{\sum_i m_i}{93.14\,h^2~{\rm eV}}~.
\label{20-omeganu}
\end{equation}
Here $\sum_i m_i$ includes all masses of the neutrino states which are
non-relativistic today. It is also useful to define the neutrino
density fraction $f_{\nu}$ with respect to the total matter density
\begin{equation}
f_{\nu} \equiv \frac{\rho_{\nu}}{(\rho_{\rm cdm}+\rho_{\rm b}
+\rho_{\nu})} =
\frac{\Omega_{\nu}}{\Omega_{\rm m}}
\end{equation}

\begin{figure}[t]
\begin{center}
\includegraphics[width=.5\textwidth]{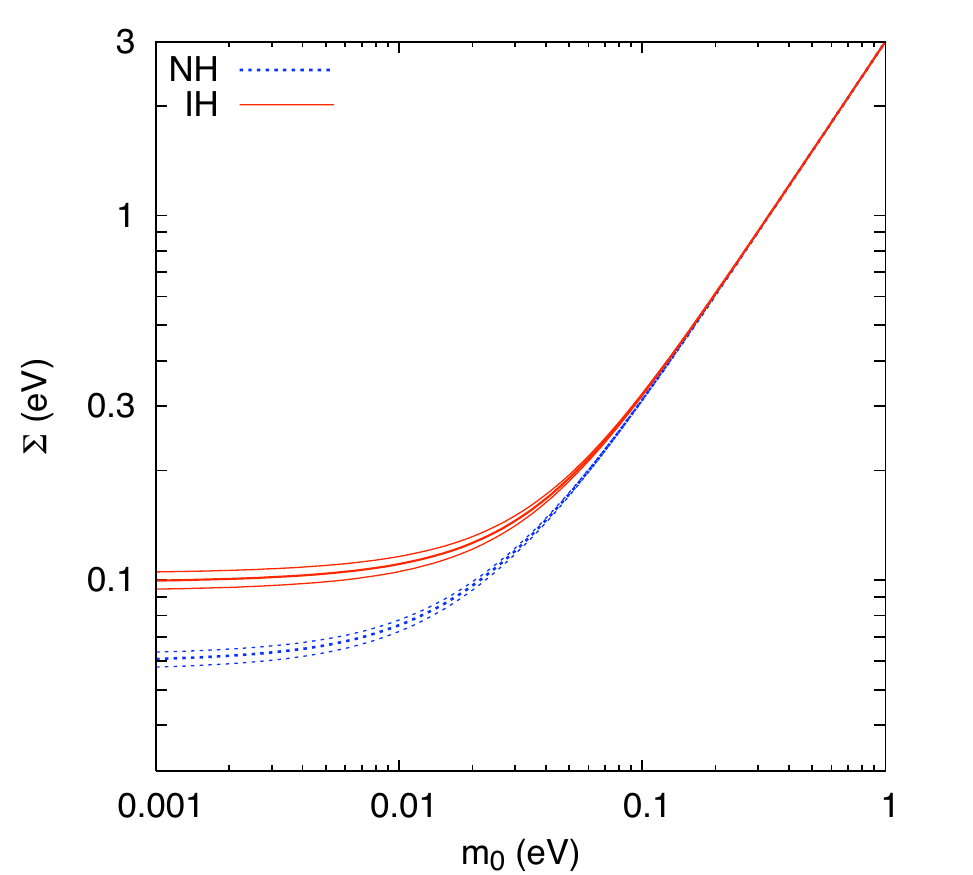}
\caption{\label{20-fig:numasses} 
Allowed values of the total neutrino mass as a
function of the lightest state within the $3\sigma$ regions of
the mixing parameters in eq.\ (\ref{20-oscpardef}). 
Blue dotted (red solid) lines correspond to normal (inverted) hierarchy for neutrino masses,
where $m_0=m_1$ ($m_0=m_3$).}
\end{center}
\end{figure}

In order to check whether relic neutrinos can have a contribution of
order unity to the present values of $\Omega_{\nu}$ or $f_{\nu}$, we
should consider which neutrino masses are allowed by non-cosmological
data. Oscillation experiments measure the differences of squared
neutrino masses $\Delta m^2_{21} = m^2_2 - m^2_1$ and $\Delta m^2_{31}
= m^2_3 - m^2_1$, the relevant ones for solar and atmospheric
neutrinos, respectively. 
As a reference, we take the
following $3\sigma$ ranges of mixing parameters from 
\cite{20-Tortola:2012te} (see also \cite{20-Fogli:2012ua,20-GonzalezGarcia:2012sz}), 
\begin{eqnarray}
\Delta m^2_{21} (10^{-5}~{\rm eV}^2)&=& 7.62_{-0.50}^{+0.58}
\nonumber \\
\Delta m^2_{31} (10^{-3}~{\rm eV}^2)& =&2.55_{-0.24}^{+0.19} \, (-2.43_{-0.22}^{+0.21})
\nonumber \\
s_{12}^2  &=&  0.32\pm{0.05}
\nonumber \\
s_{23}^2  &\in& [0.36,0.68]\, ([0.37,0.67])\nonumber \\
s_{13}^2  &=&  0.0246_{-0.0076}^{+0.0084}\, (0.025 \pm 0.008) 
\label{20-oscpardef}
\end{eqnarray}
Here $s^2_{ij}=\sin^2 \theta_{ij}$, where $\theta_{ij}$ ($ij=12, 23$ or
$13$) are the three mixing angles.  Unfortunately oscillation
experiments are insensitive to the absolute scale of neutrino masses,
because the knowledge of $\Delta m^2_{21}>0$ and $|\Delta m^2_{31}|$
leads to the two possible schemes shown in fig.\ 1 of
\cite{20-Lesgourgues:2006nd}, but leaves one neutrino mass unconstrained. These
two schemes are known as normal (NH) and inverted (IH) mass hierarchies,
characterized by the sign of $\Delta m^2_{31}$, positive and negative,
respectively.  In the above equation the values in parentheses 
correspond to the IH, otherwise the mixing parameters present the same allowed regions for both hierarchies. For small values of the lightest neutrino mass $m_0$,
i.e.\ $m_1$ ($m_3$) for NH (IH), the mass states follow a hierarchical
scenario, while for masses much larger than the differences all
neutrinos share in practice the same mass and then we say that they
are degenerate. In general, the relation between the individual masses
and the total neutrino mass can be found numerically, as shown in
Fig.\ \ref{20-fig:numasses}.

\begin{figure}[t]
\begin{center}
\hspace{-1.cm}
\includegraphics[width=.52\textwidth]{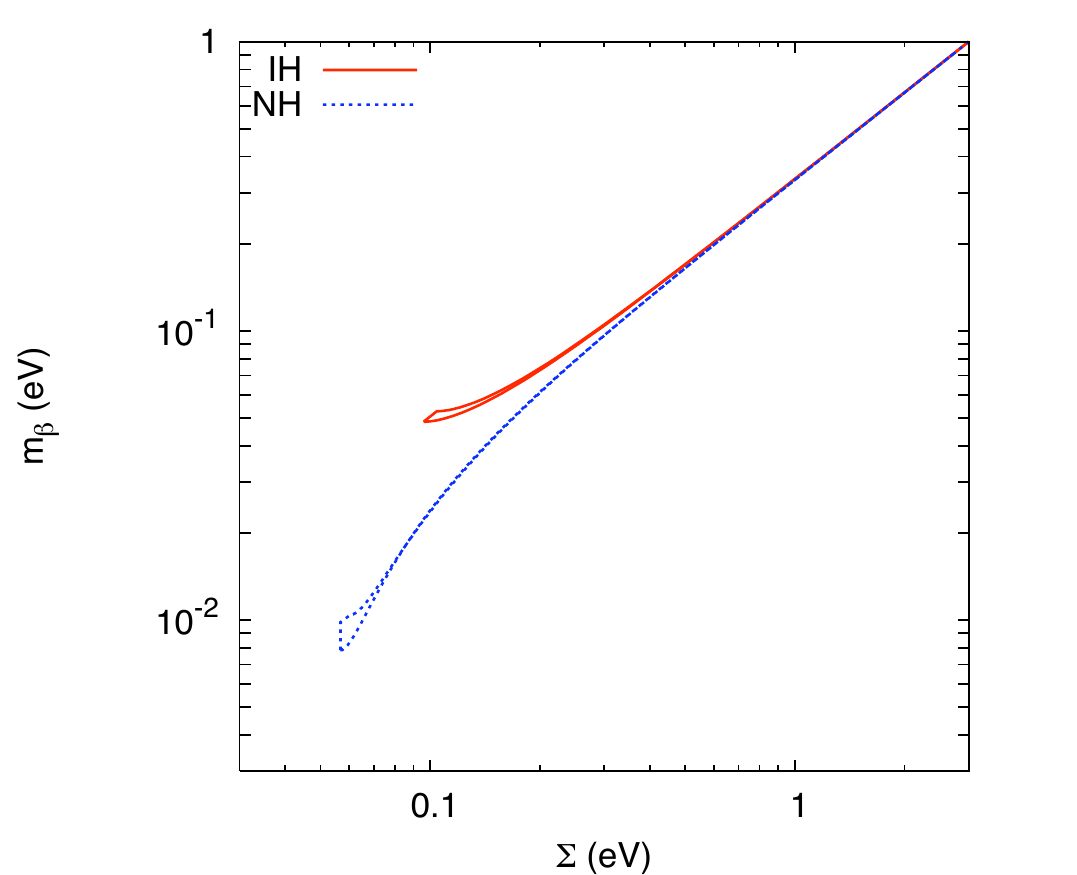}
\hspace{-0.5cm}
\includegraphics[width=.52\textwidth]{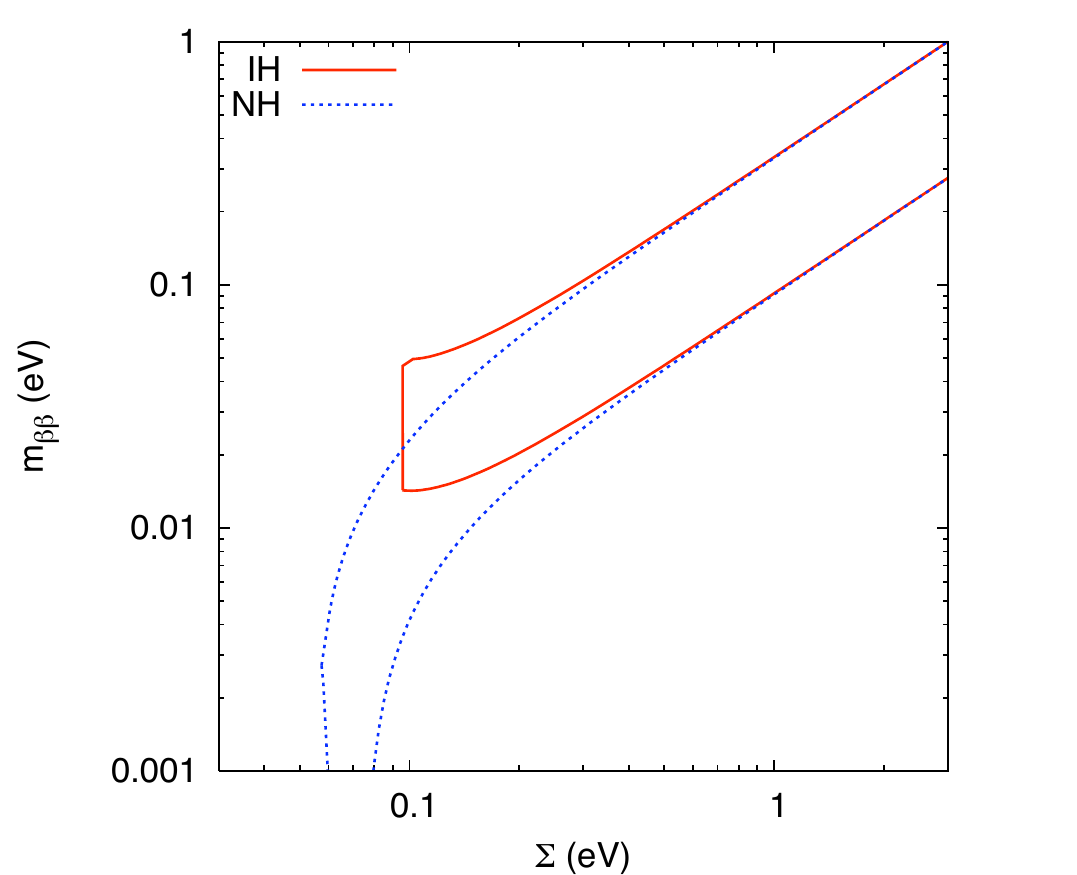}
\caption{\label{20-fig:mee} Allowed regions by oscillation data at the 3$\sigma$ level (eq.\ (\ref{20-oscpardef}))
of the three main observables sensitive to the absolute scale of neutrino masses. We show the regions
in the planes $m_{\beta}-\Sigma$ and $m_{\beta\beta}-\Sigma$ , see the text for the corresponding definitions.
Blue dotted (red solid) regions correspond to normal (inverted) hierarchy.}
\end{center}
\end{figure}

There are two types of laboratory experiments searching for the
absolute scale of neutrino masses, a crucial piece of information for
constructing models of neutrino masses and mixings.
The neutrinoless double beta decay $(Z,A) \to (Z+2,A)+2e^-$ (in short
$0\nu2\beta$) is a rare nuclear processes where lepton number is
violated and whose observation would mean that neutrinos are Majorana
particles. If the $0\nu2\beta$ process is mediated by a light
neutrino, the results from neutrinoless double beta decay experiments
are converted into an upper bound or a measurement of the effective
mass $m_{\beta\beta}$
\begin{equation}
m_{\beta\beta} 
= |c_{12}^2c_{13}^2\, m_1+s_{12}^2c_{13}^2\, m_2\, e^{i\phi_2}
+s_{13}^2\, m_3\, e^{i\phi_3}|~,
\label{20-beta2beta}
\end{equation}
where $\phi_{1,2}$ are the two Majorana phases that appear in
lepton-number-violating processes. An important issue for $0\nu2\beta$
results is related to the uncertainties on the corresponding nuclear
matrix elements. For more details
and the current experimental results, see \cite{20-Giuliani}.

Beta decay experiments, which involve only the kinematics of
electrons, are in principle the best strategy for measuring directly
the neutrino mass \cite{20-Drexlin}.  The current limits from tritium
beta decay apply only to the range of degenerate neutrino masses, so
that $m_\beta \simeq m_0$, where
\begin{equation}
m_\beta 
= (c_{12}^2c_{13}^2\, m_1^2+s_{12}^2c_{13}^2\, m_2^2
+s_{13}^2\, m_3^2)^{1/2},
\label{20-beta}
\end{equation}
is the relevant parameter for beta decay experiments. The bound at
$95\%$ CL is $m_0<2.05-2.3$ eV from the Troitsk and Mainz experiments,
respectively. This value is expected to be improved by the KATRIN
project to reach a discovery potential for $0.3-0.35$ eV masses (or a
sensitivity of $0.2$ eV at $90\%$ CL). Taking into account this upper
bound and the minimal values of the total neutrino mass in the normal
(inverted) hierarchy, the sum of neutrino masses is restricted to the
approximate range
\begin{equation}
0.06\, (0.1)~{\rm eV} \lesssim \sum_i m_i \lesssim 6~{\rm eV}
\label{20-sum_range}
\end{equation}

As we discuss in the next sections, cosmology is at first order
sensitive to the total neutrino mass $\sum\equiv\sum_i m_i$ if all states have the same
number density, providing information on $m_0$ but blind to neutrino
mixing angles or possible CP violating phases. Thus cosmological
results are complementary to terrestrial experiments. The interested
reader can find the allowed regions in the parameter space defined by
any pair of parameters $(\sum,m_{\beta\beta},m_\beta)$ in
\cite{20-Fogli:2005cq,20-Fogli:2008ig,20-GonzalezGarcia:2010un}.
The two cases involving $\sum$ are shown in Fig.\ \ref{20-fig:mee}.

Now we can find the possible present values of $\Omega_\nu$ in
agreement  the approximate bounds of eq.\ (\ref{20-sum_range}).
Note that even if the three neutrinos are non-degenerate in mass, eq.\
(\ref{20-omeganu}) can be safely applied, because we know from neutrino
oscillation data that at least two of the neutrino states are
non-relativistic today, because both $(\Delta m^2_{31})^{1/2}\simeq 0.05
$ eV and $(\Delta m^2_{21})^{1/2}\simeq 0.009$ eV are larger than the
temperature $T_\nu \simeq 1.96$ K $\simeq 1.7\times 10^ {-4}$ eV. If
the third neutrino state is very light and still relativistic, its
relative contribution to $\Omega_{\nu}$ is negligible and eq.\
(\ref{20-omeganu}) remains an excellent approximation of the total
density. One finds that $\Omega_\nu$ is restricted to the approximate
range
\begin{equation}
0.0013\, (0.0022) \lesssim \Omega_{\nu} \lesssim 0.13
\label{20-omega_nu_limits}
\end{equation}
where we already included that $h\approx 0.7$. This applies only to
the standard case of three light active neutrinos, while in general a
cosmological upper bound on $\Omega_{\nu}$ has been used since the
1970s to constrain the possible values of neutrino masses. For
instance, if we demand that neutrinos should not be heavy enough to
overclose the Universe ($\Omega_{\nu}<1$), we obtain an upper bound
$\sum \lesssim 45$ eV (again fixing $h=0.7$). Moreover, because from
present analyses of cosmological data we know that the approximate
contribution of matter is $\Omega_{\rm m} \simeq 0.3$, the neutrino
masses should obey the stronger bound $\sum \lesssim 15$ eV. We see
that with this simple argument one obtains a bound which is roughly
only a factor 2 worse than the bound from tritium beta decay, but of
course with the caveats that apply to any cosmological analysis.  In
the three-neutrino case, these bounds should be understood in terms of
$m_0=\sum/3$.

Dark matter particles with a large velocity dispersion such as that of
neutrinos are called hot dark matter (HDM). The role of neutrinos as
HDM particles has been widely discussed since the 1970s, and the
reader can find a historical review in \cite{20-Primack:2001ib}.
It was realized in the mid-1980s that HDM affects the evolution of
cosmological perturbations in a particular way: it erases the density
contrasts on wavelengths smaller than a mass-dependent free-streaming
scale. In a universe dominated by HDM, this suppression is in
contradiction with various observations. For instance, large objects
such as superclusters of galaxies form first, while smaller structures
like clusters and galaxies form via a fragmentation process. This
top-down scenario is at odds with the fact that galaxies seem older
than clusters.

Given the failure of HDM-dominated scenarios, the attention then
turned to cold dark matter (CDM) candidates, i.e.\ particles which
were non-relativistic at the epoch when the universe became
matter-dominated, which provided a better agreement with
observations. Still in the mid-1990s it appeared that a small mixture
of HDM in a universe dominated by CDM fitted better the observational
data on density fluctuations at small scales than a pure CDM
model. However, within the presently favoured $\Lambda$CDM model
dominated at late times by a cosmological constant (or some form of
dark energy) there is no need for a significant contribution of
HDM. Instead, one can use the available cosmological data to find how
large the neutrino contribution can be, as we will see later.

Before concluding this section, we would like to mention the case of a
sterile neutrino with a mass of the order of a few keV's and a very
small mixing with the flavour neutrinos. Such ``heavy'' neutrinos
could be produced by active-sterile oscillations but not fully
thermalized, so that they could play the role of dark matter and
replace the usual CDM component. But due to their large thermal
velocity (slightly smaller than that of active neutrinos), they would
behave as Warm Dark Matter and erase small-scale cosmological
structures.  Their mass can be bounded from below using Lyman-$\alpha$
forest data from quasar spectra, and from above using X-ray
observations. The viability of this scenario is currently under
careful examination (see e.g.\ \cite{20-Boyarsky:2009ix}).

\section{Effects of neutrino masses on cosmology}
\label{20-sec:effects_numass}

In this section we will briefly describe the main cosmological
observables and the effects that neutrino masses cause on them.  A
more detailed discussion of the effects of massive neutrinos on the
evolution of cosmological perturbations can be found in Secs.\ 4.5
and 4.6 of \cite{20-Lesgourgues:2006nd}.

\subsection{Brief description of cosmological observables}

Although there exist many different types of cosmological
measurements, here we will restrict the discussion to those that are
at present the more important for obtaining an upper bound or
eventually a measurement of neutrino masses.

First of all, we have the CMB temperature anisotropy power spectrum, defined as the angular two-point correlation
function of CMB maps $\delta T/\bar{T}(\hat{n})$ ($\hat{n}$ being a
direction in the sky). This function is usually expanded in Legendre
multipoles
\begin{equation}
\left\langle {\delta T\over\bar{T}}(\hat{n}) 
{\delta T \over \bar{T}} (\hat{n}') \right\rangle
= \sum_{l=0}^{\infty} {(2l+1)\over4\pi} \,C_l\, P_l(\hat{n}\cdot\hat{n}')~,
\end{equation}
where $P_l(x)$ are the Legendre polynomials.  For Gaussian
fluctuations, all the information is encoded in the multipoles $C_l$
which probe correlations on angular scales $\theta=\pi/l$.  We have
seen that each neutrino state can only have a mass of the order of
$1$ eV, so that the transition of relic neutrinos to the
non-relativistic regime is expected to take place after the time of
recombination between electrons and nucleons, i.e.\ after photon
decoupling. Because the shape of the CMB spectrum is related mainly to
the physical evolution {\it before} recombination, it will only be
marginally affected by the neutrino mass, except through a modified background evolution and some secondary anisotropy corrections. There exists
interesting complementary information to the temperature power
spectrum if the CMB polarization is measured, and currently we have
some less precise data on the temperature $\times$ E-polarization (TE)
correlation function and the E-polarization self-correlation spectrum
(EE).

The current Large Scale Structure (LSS) of the Universe is probed by
the matter power spectrum, observed with various techniques described
in the next section (directly or indirectly, today or in the near past
at redshift $z$). It is defined as the two-point correlation function
of non-relativistic matter fluctuations in Fourier space
\begin{equation}
P(k,z)=\langle | \delta_{\rm m}(k,z) |^2 \rangle~,
\end{equation}
where $\delta_{\rm m}=\delta \rho_{\rm m}/\bar{\rho}_{\rm m}$.
Usually $P(k)$ refers to the matter power spectrum evaluated today (at
$z=0$).  In the case of several fluids (e.g.\ CDM, baryons and
non-relativistic neutrinos), the total matter perturbation can be
expanded as
\begin{equation}
\delta_{\rm m}= 
\frac{\sum_i \, \bar{\rho}_i \, \delta_i}{\sum_i \, \bar{\rho}_i}~.
\end{equation}
Because the energy density is related to the mass density of
non-relativistic matter through $E=mc^2$, $\delta_{\rm m}$ represents
indifferently the energy or mass power spectrum. The shape of the
matter power spectrum is affected in a scale-dependent way by the
free-streaming caused by small neutrino masses of ${\cal O}$(eV) and
thus it is the key observable for constraining $m_\nu$ with
cosmological methods.

\subsection{Neutrino free-streaming}

After thermal decoupling, relic neutrinos constitute a collisionless
fluid, where the individual particles free-stream with a
characteristic velocity that, in average, is the thermal velocity
$v_{\rm th}$. It is possible to define a horizon as the typical
distance on which particles travel between time $t_i$ and $t$.  When
the Universe was dominated by radiation or matter $t \gg t_i$, this
horizon is, as usual, asymptotically equal to $v_{\rm th}/H$, up to a
numerical factor of order one. Similar to the definition of the Jeans
length (see section 4.4 in \cite{20-Lesgourgues:2006nd}), we can define the
neutrino free-streaming wavenumber and length as
\begin{eqnarray}
k_{FS}(t) &=& \left(\frac{4 \pi G \bar{\rho}(t) 
a^2(t)}{v_{\rm th}^2(t)}\right)^{1/2}, \\
\lambda_{FS}(t) 
&=& 2 \pi \frac{a(t)}{k_{FS}(t)}
= 2 \pi \sqrt{2 \over 3} \frac{v_{\rm th}(t)}{H(t)}~.
\end{eqnarray}
%
As long as neutrinos are relativistic, they travel at the speed of
light and their free-streaming length is simply equal to the Hubble
radius. When they become non-relativistic, their thermal velocity
decays like
\begin{equation}
v_{\rm th}\equiv\frac{\langle p \rangle}{m}
\simeq\frac{3.15 T_{\nu}}{m}=\frac{3.15 T_{\nu}^0}{m}
\left( \frac{a_0}{a} \right) 
\simeq 158 (1+z) \left( \frac{1 \, \mathrm{eV}}{m} \right)
{\rm km}\,{\rm s}^{-1}~,
\end{equation}
where we used for the present neutrino temperature $T_{\nu}^0 \simeq
(4/11)^{1/3} T_{\gamma}^0$ and $T_{\gamma}^0 \simeq 2.726$ K. This
gives for the free-streaming wavelength and wavenumber during matter
or ${\Lambda}$ domination
\begin{eqnarray}
k_{FS}(t) &=& 0.8
\frac{\sqrt{\Omega_{\Lambda}+ \Omega_{m} (1+z)^3}}{(1+z)^2}
\left( \frac{m}{1 \, \mathrm{eV}} \right) h\,\mathrm{Mpc}^{-1}, \\
\lambda_{FS}(t) 
&=& 
8 
\frac{1+z}{\sqrt{\Omega_{\Lambda}+ \Omega_{m} (1+z)^3}}
\left( \frac{1 \, \mathrm{eV}}{m} \right) h^{-1}\mathrm{Mpc}~,
\end{eqnarray}
where $\Omega_{\Lambda}$ and $\Omega_{m}$ are the cosmological
constant and matter density fractions, respectively, evaluated today.
After the non-relativistic transition and during matter
domination, the free-streaming length continues to increase, but only
like $(aH)^{-1}\propto t^{1/3}$, i.e.\ more slowly than the scale
factor $a \propto t^{2/3}$.  Therefore, the comoving free-streaming
length $\lambda_{FS} / a$ actually decreases like $(a^2 H)^{-1}
\propto t^{-1/3}$. As a consequence, for neutrinos becoming
non-relativistic during matter domination, the comoving free-streaming
wavenumber passes through a minimum $k_{\rm nr}$ at the time of the
transition, i.e.\ when $m = \langle p \rangle = 3.15 T_{\nu}$ and
$a_0/a=(1+z)= 2.0\times 10^3 (m/ 1 \, \mathrm{eV})$. This minimum
value is found to be
\begin{equation}
k_{\rm nr}
\simeq 0.018 \,\, \Omega_{\rm m}^{1/2} 
\left( \frac{m}{1 \, \mathrm{eV}} \right)^{1/2}
h\,\mathrm{Mpc}^{-1}~.
\label{20-knr}
\end{equation}
The physical effect of free-streaming is to damp small-scale neutrino
density fluctuations: neutrinos cannot be confined into (or kept
outside of) regions smaller than the free-streaming length, because their velocity is greater than the escape velocity from gravitational potential wells on those scales.
Instead, on scales much larger than the free-streaming scale, the
neutrino velocity can be effectively considered as vanishing, and after
the non-relativistic transition the neutrino perturbations behave like
CDM perturbations. In particular, modes with $k<k_{\rm nr}$ are never
affected by free-streaming and evolve like being in a pure $\Lambda$CDM
model.

\subsection{Impact of massive neutrinos on the matter power spectrum}
\label{20-subsec:impact_pk}

The small initial cosmological perturbations evolve within the linear regime at early times. During matter domination, the smallest cosmological scales start evolving non-linearily, leading to the formation of the structures we see today. In the recent universe, the largest observable scales are still related to the linear evolution, while other scales can only be understood using non-linear N-body simulations. We will not review here all the details of this complicated evolution 
(see \cite{20-NuCosmo,20-Wong:2011ip,20-Lesgourgues:2006nd} and
references therein), but we will emphasize the main effects caused by
massive neutrinos on linear scales in the framework of the standard cosmological
scenario: a $\Lambda$ Mixed Dark Matter ($\Lambda$MDM) model, where
Mixed refers to the inclusion of some HDM component.

On large scales (i.e. on wave-numbers smaller than the value $k_{\rm nr}$ defined in the previous subsection), neutrino free-streaming can be ignored, and neutrino perturbations are indistinguishable from CDM perturbations. On those scales, the matter power spectrum $P(k,z)$ can be shown to depend only on the matter density fraction today (including neutrinos), $\Omega_{\rm m}$, and on the primordial perturbation spectrum. If the neutrino mass is varied with $\Omega_{\rm m}$ fixed, the large -scale power spectrum remains invariant.

On small scales such that $k>k_{\rm nr}$, the matter power spectrum is affected by neutrino masses for essentially three reasons:
\begin{enumerate}
\item massive neutrinos do not cluster on those scales. The matter power spectrum can be expanded as a function of the three non-relativistic species,
\begin{equation}
P(k,z) = \left\langle 
\left| \frac{\delta \rho_{\rm cdm} + \delta \rho_{\rm b} + \delta \rho_{\nu}}
{\rho_{\rm cdm} + \rho_{\rm b} + \rho_{\nu}}\right|^2 \right\rangle 
=  \Omega_{\rm m}^{-2} \left\langle \left| 
\Omega_{\rm cdm} \, \delta_{\rm cdm} 
+ \Omega_{\rm b} \, \delta_{\rm b} 
+ \Omega_{\nu} \, \delta_{\nu}
\right|^2
\right\rangle~.
\end{equation}
On scales of interest and in the recent universe, baryon and CDM fluctuations are equal to each other, while $\delta_\nu \ll \delta_{\rm cdm}$. Hence, even if the evolution of $ \delta_{\rm cdm}$ was not affected by neutrino masses (which is not the case, see the remaining two points below), the power spectrum would be reduced by a factor $(1-f_\nu)^2$ with
\begin{equation}
f_\nu \equiv \frac{\Omega_\nu}{\Omega_{\rm m}}.
\end{equation}
\item
the redshift of radiation-to-matter equality $z_{\rm eq}$ or the baryon-to-CDM ratio $\omega_{\rm b}/\omega_{\rm cdm}$ might be slightly affected by neutrino masses, with a potential impact on the small-scale matter power spectrum. This depends very much on which other parameters are kept fixed when the neutrino mass varies. If neutrino masses are smaller than roughly 0.5~eV, they are still relativistic at the time of radiation-to-matter equality, and the redshift of equality depends on $(\omega_{\rm b}+\omega_{\rm cdm})$, not on $\omega_{\rm m}$. It is possible to increase $M_\nu$ and $\omega_{\rm m}$ with fixed parameters $\Omega_{\rm m}$, $\omega_{\rm b}$, $\omega_{\rm cdm}$ (provided that the Hubble parameter also increases like the square root of $\omega_{\rm m}$). In that case, there is no significant impact of massive neutrinos on the matter power spectrum through background effects, i.e. through a change in the homogeneous cosmological evolution. However, there are some important perturbation effects that we will now summarize.
\item
the growth rate of cold dark matter perturbations is reduced through an absence of gravitational back-reaction effects from free-streaming neutrinos. This growth rate is set by an equation of the type
\begin{equation}
\delta_{\rm cdm}'' + \frac{a'}{a} \delta_{\rm cdm} = -k^2 \psi~,
\end{equation}
where $\delta_{\rm cdm}$ stands for the CDM relative density perturbation in Fourier space, and $\psi$ for the metric perturbation playing the role of the Newtonian potential inside the Hubble radius. There is a similar equation for decoupled baryons, and very soon after baryon decoupling we can identify $\delta_{\rm b} = \delta_{\rm cdm}$ on scales of interest. On the right-hand side, we neglected time derivatives of metric fluctuations playing only a minor role. The right-hand side represents gravitational clustering, and is given by the Poisson equation as a function of the total density fluctuation. The second term on the left-hand side represents Hubble friction, i.e. the fact that the cosmological expansion enhances distances, reduces gravitational forces and slows down gravitational clustering. The coefficient $a'/a$ is given by the Friedmann equation as a function of the total background density. In a universe such that all species present in the Friedmann equation do cluster, as it is the case in a matter-dominated universe with $\delta \rho_{\rm total} \simeq \delta \rho_{\rm cdm} + \delta \rho_b$ and $\bar{\rho}_{\rm total} = \bar{\rho}_{\rm cdm} + \bar{\rho}_{\rm b}$, the solution is simply given by $\delta_{\rm cdm} \propto a$: the so-called linear growth factor is proportional to the  scale factor. But whenever one of the species contributing to the background expansion does not cluster efficiently, it can be neglected in the Poisson equation. In that case, the term on the right-hand side becomes smaller with respect to the Hubble friction term, and CDM (as well as baryons) clusters at a slower rate. This is the case in presence of massive neutrinos and for $k \gg k_{\rm nr}$: the linear growth rate during matter domination is then equal to $a^{1-3/5 f_\nu}$.  During $\Lambda$ domination, this effect sums up with that of the cosmological constant (or of any non-clustering dark energy), which also tends to reduce the growth rate for the very same reason.
\end{enumerate}

In summary, the small-scale matter power spectrum $P(k \geq k_{\rm nr})$ is reduced in presence of massive neutrinos for at least two reason: by the absence of neutrino perturbations in the total matter power spectrum, and by a slower growth rate of CDM/baryon perturbations at late times. The third effect has the largest amplitude. At low redhsift $z \simeq 0$, the step-like suppression of $P(k)$ starts at $k \geq k_{\rm nr}$ and saturates at $k \sim 1 h/$Mpc with a constant amplitude $\Delta P(k)/P(k) \simeq -8 f_\nu$. This result was obtained by fitting numerical simulations \cite{20-Hu:1997mj}, but a more accurate approximation can be derived analytically 
\cite{20-NuCosmo,20-Lesgourgues:2006nd}. As mentioned in the second item above, neutrino masses can have additional indirect effects through a change in the background evolution, depending on which cosmological parameters are kept fixed when $M_\nu$ varies.

\begin{figure}[t]
\begin{center}
\vspace{-6cm}
\includegraphics[width=0.65\textwidth]{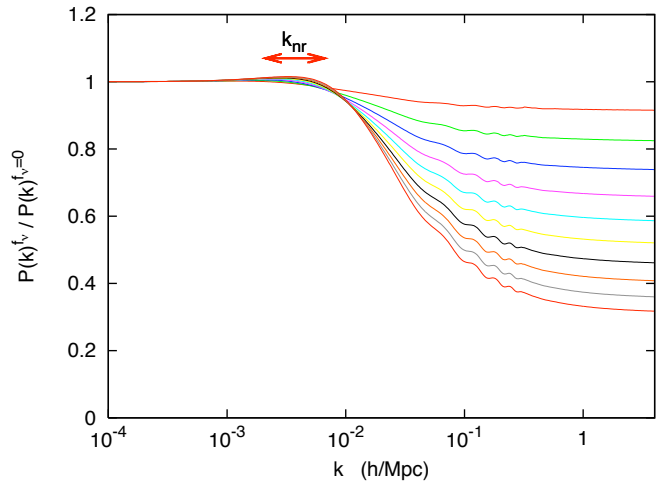}\\
\caption{\label{20-fig_tk} Ratio of the matter power spectrum including
three degenerate massive neutrinos with density fraction $f_{\nu}$ to
that with three massless neutrinos.  The parameters $(\omega_{\rm m},
\, \Omega_{\Lambda})=(0.147,0.70)$\ are kept fixed, and from top to
bottom the curves correspond to $f_{\nu}=0.01, 0.02, 0.03,\ldots,0.10$.
The individual masses $m_{\nu}$ range from $0.046$ to $0.46$ eV,
and the scale $k_{\rm nr}$ from $2.1\times10^{-3}h\,$Mpc$^{-1}$ to
$6.7\times10^{-3}h\,$Mpc$^{-1}$ as shown on the top of the
figure. {}From \cite{20-Lesgourgues:2006nd}.}
\end{center}
\end{figure}

When fitting data, one can use
analytical approximations to the full MDM or $\Lambda$MDM matter power
spectrum, valid for arbitrary scales and redshifts, as listed in
\cite{20-Lesgourgues:2006nd}. However, nowadays the analyses are performed
using the matter power spectra calculated by Boltzmann codes such as 
{\sc camb} \cite{20-Lewis:1999bs} or {\sc class} \cite{20-Lesgourgues:2011rh},
that solve numerically the evolution of the cosmological
perturbations.  The step-like suppression of the matter power spectrum induced by various values of $f_\nu$ is shown in Fig.\ \ref{20-fig_tk}.

Is it possible to mimic the effect of massive neutrinos on the matter
power spectrum with some combination of other cosmological parameters?
If so, one would say that a parameter degeneracy exists, reducing the
sensitivity to neutrino masses. This possibility depends on the
intervals $[k_{\rm min},k_{\rm max}]$ and $[z_{\rm min}, z_{\rm max}]$ in which $P(k,z)$ can be
accurately measured. Ideally, if we could have $k_{\rm min} \leq
10^{-2}h\,$Mpc$^{-1}$ and $k_{\rm max} \geq 1\,h\,$Mpc$^{-1}$, the
effect of the neutrino mass would be non-degenerate, because of its
very characteristic step-like effect. Moreover, because neutrinos render the linear growth factor scale-dependent, with
$\delta_{\rm m} (k,z)$ proportional to $a$ (resp. or $a^{1-3/5f_\nu}$) for $k < k_{\rm nr}$ (resp. $k \gg k_{\rm nr})$,
the amplitude of the step-like suppression is redshift dependent. Using e.g. weak lensing techniques or Lyman-$\alpha$ forest data, one could get accurate measurements of the matter spectrum at  at several redshifts in the range $0<z<3$. This will offer an opportunity to test the ``time dependence'' of the neutrino mass effect, and to distinguish it, for instance, from an equivalent step-like suppression in the primordial spectrum, that would still imply a scale-independent growth factor.

The problem is that usually the matter power spectrum can only be
accurately measured in the intermediate region where the mass effect
is neither null nor maximal: in other words, many experiments only
have access to the transition region in the step-like transfer
function. In this region, the neutrino mass affects the slope of the
matter power spectrum in a way which can be easily confused with the
effect of other cosmological parameters. Because of these parameter
degeneracies, the current LSS data alone cannot provide significant
constraints on the neutrino mass, and it is necessary to combine them
with other cosmological data, in particular the CMB anisotropy
spectrum, which could lift most of the degeneracies.  Still, for
exotic models with e.g.\ extra relativistic degrees of freedom, a
constant equation-of-state parameter of the dark energy different from
$-1$ or a non-power-law primordial spectrum, the
neutrino mass bound can become significantly weaker.

The value of $k_{\rm max}$ is not limited by observational sensitivities, but by the range in which we trust predictions from linear theory. Beyond the scale of non-linearity (of the order of $k_{\rm max} = 0.15 h\,$Mpc$^{-1}$ at $z=0$), the data is only useful provided that one is able to make accurate predictions not only for the non-linear power spectrum, but also for redshift-space distortions (coming from the fact that we observe the redshift of objects, not their distance away from us), and finally for the scale dependence of the light-to-mass bias (relating the  power spectrum of a given category of observed compact objects to the underlying total matter power spectrum). Spectacular progresses are being carried on these three fronts, thanks to better N-body simulations and analytical techniques. Including massive neutrinos in such calculations is particularly difficult, but successful attempts were presented e.g. in \cite{20-Brandbyge:2009ce,20-Bird:2011rb,20-Wagner:2012sw}. The neutrino mass impact on the non-linear matter power spectrum is now modeled with rather good precision, at least  within one decade above $k_{\rm max}$ in wave-number space. It appears that the step-like suppression is enhanced by non-linear effects up to roughly $\Delta P(k)/P(k) \simeq -10 f_\nu$ (at redshift zero and $k\sim 1\,h\,$Mpc$^{-1}$), and is reduced above this scale. Hence, non-linear corrections render the neutrino mass effect even more characteristic than in the linear theory, and may help to increase the sensitivity of future experiments.

Until this point, we reduced the neutrino mass effect to that of the parameter $f_\nu$ or $M_\nu$. In principle, the mass splitting between the three different families for a common total mass is visible in the matter power spectrum. The time at which each species becomes non-relativistic depends on individual masses $m_i$. Hence, both the scale of the step-like suppression induced by {\it each} neutrino and the amount of suppression in the small-scale power spectrum have a small dependence on individual masses. The differences between the power spectrum of various models with the same total mass and different mass splittings was computed numerically in \cite{20-Lesgourgues:2004ps} for the linear spectrum, and \cite{20-Wagner:2012sw} for the non-linear spectrum. At the moment, it seems that even the most ambitious future surveys will not be able to distinguish these mass splitting effects with a reasonable significance \cite{20-Jimenez:2010ev,20-Pritchard:2008wy}.

\subsection{Impact of massive neutrinos on the CMB anisotropy spectrum}

For neutrino masses of the order of $1$ eV (about $f_{\nu} \leq 0.1$)
the three neutrino species are still relativistic at the time of
photon decoupling, and the direct effect of free-streaming neutrinos
on the evolution of the baryon-photon acoustic oscillations is the
same in the $\Lambda$CDM and $\Lambda$MDM cases. Therefore, the effect
of the mass can only appear at two levels: that of the background
evolution, and that of secondary CMB anisotropies, related to the behavior of photon perturbations after decoupling. Both levels are potentially affected by the evolution of neutrinos after the time of their non-relativistic transition. If neutrinos were heavier than a few eV, they would
already be non-relativistic at decoupling, and they could trigger more direct effects in the CMB, as described in \cite{20-Dodelson:1995es}.  However, we will see later that this
situation is disfavoured by current upper bounds on the neutrino mass.

Let us first review the background effects of massive neutrinos on  the CMB. Because the temperature and polarization spectrum shape is the result of several intricate effects, one cannot discuss the neutrino mass impact without specifying which other parameters are kept fixed. Neutrinos with a mass in the range from $10^{-3}$ eV to $1$ eV should be counted as radiation at the time of equality, and as non-relativistic matter today: the total non-relativistic density, parametrized by $\omega_{\rm m}=\Omega_{\rm m} h^2$, depends on the total neutrino mass $M_\nu = \sum_i m_i$. Hence, when $M_\nu$ is varied, there must be a variation either in the redshift of matter-to-radiation equality $z_{\rm eq}$, or in the matter density today $\omega_{\rm m}$.

This can potentially impact the CMB in three ways. A shift in the redshift of equality affects the position and amplitude of the peaks. A change in the non-relativistic matter density at late times can impact both the angular diameter distance to the last scattering surface $d_A(z_{\rm dec})$, controlling the overall position of CMB spectrum features in multipole space, and the slope of the low-$l$ tail of the CMB spectrum, due to the late Integrated Sachs-Wolfe (ISW) effect. Out of these three effects (changes in $z_{\rm eq}$, in $d_A$ and in the late ISW), only two can be cancelled by a simultaneous variation of the total neutrino mass and of other free parameters in the $\Lambda$MDM model. Hence, the CMB spectrum is sensitive to the background effect of the total neutrino mass. In practice however, the late ISW effect is difficult to measure due to cosmic variance and CMB data alone cannot provide a useful information on sub-eV neutrino masses. If one considers extensions of the $\Lambda$MDM, this becomes even more clear: by playing with the spatial curvature, one can neutralize all three effects simultaneously. But as soon as CMB data is used in combination with other background cosmology observations (constraining for instance the Hubble parameter, the cosmological constant value or the BAO scale), some bounds can be derived on $M_\nu$.

There exists another effect of massive neutrinos on the CMB at the level of secondary anisotropies: when neutrinos become non-relativistic, they  reduce the time variation  of the gravitational potential inside the Hubble radius. This affects the photon temperature through the early ISW effect, and leads to a depletion in the temperature spectrum of the order of $(\Delta C_l/C_l) \sim -(M_\nu/0.1 \, {\rm eV}) \%$  on multipoles $20<l<500$, with a dependence of the maximum depletion scale on individual masses $m_i$. This effect is roughly ten times smaller than the depletion in the small-scale matter power spectrum, $\Delta P(k)/P(k) \sim -(M_\nu/0.01 \, {\rm eV}) \%$. 

\begin{figure}[t]
\begin{center}
\includegraphics[width=0.65\textwidth]{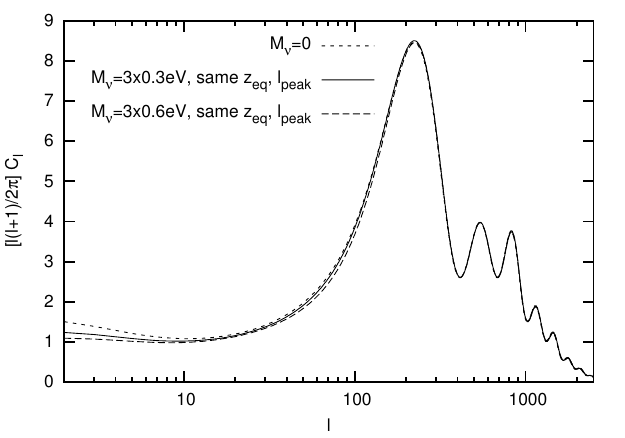}\\
\caption{\label{20-fig_cmb} CMB temperature spectrum with different neutrino masses. Some of the parameters of the $\Lambda$MDM model have been varied together with $M_\nu$ in order to keep fixed the redshift of equality and the angular diameter distance to last scattering.}
\end{center}
\end{figure}

We show in Figure~\ref{20-fig_cmb}  the effect on the CMB temperature spectrum of increasing the neutrino mass while keeping $z_{\rm eq}$ and $d_A$ fixed: the only observed differences are then for $2<l<50$ (late ISW effect due to neutrino background evolution) and for $50<l<200$ (early ISW effect due to neutrino perturbations).

We conclude that the CMB alone is not a very powerful tool for constraining sub-eV neutrino masses, and should be used in combination with homogeneous cosmology constraints and/or measurements of the LSS power spectrum, for instance from  galaxy clustering, galaxy lensing or CMB lensing (see section \ref{20-sec:future}).

\section{Current bounds on neutrino masses}
\label{20-sec:current}

In this section we review how the available cosmological data are used
to get information on the absolute scale of neutrino masses,
complementary to laboratory experiments. Note that the bounds in the
next subsections are all based on the Bayesian inference method, and
the upper bounds on the sum of neutrino masses are given at 95\% C.L.\
after marginalization over all free cosmological parameters. We refer
the reader to section 5.1 of \cite{20-Lesgourgues:2006nd} for a detailed
discussion on this statistical method, as well as for most of the
references for the experimental data or parameter analysis.

\subsection{CMB anisotropy bounds}
\label{20-sec:present_CMB}

The experimental situation of the measurement of the CMB anisotropies
is dominated by the seven-year release of WMAP data \cite{20-Komatsu:2010fb}, which 
improved the already precise TT and TE
angular power spectra of the previous releases, and
included a detection of the E-polarization self-correlation spectrum (EE).
On similar or smaller angular scales than WMAP, we have results from
experiments that are either ground-based or balloon-borne (ACT, SPT, \ldots).  

We saw in the previous section that the CMB spectrum has a small sensitivity to neutrino masses even when each mass is below 0.5~eV, and all non-relativistic transitions take place after photon decoupling 
\cite{20-Ichikawa:2004zi}. This sensitivity is due to background effects (mainly  the late ISW effect if all other $\Lambda$MDM parameters are left free) plus perturbation effects (mainly the early ISW effect). Therefore, it is possible to
constrain neutrino masses using CMB experiments only, down to the
level at which these small effects are masked by instrumental noise,
or by cosmic variance, or by parameter degeneracies in the case of
some cosmological models beyond the minimal $\Lambda$ Mixed Dark
matter framework. WMAP alone is able to set a limit
$M_\nu<1.3-1.4$ eV
 depending on whether
the dark energy component is a cosmological constant or not \cite{20-Komatsu:2010fb}. Because the neutrino mass effects in the CMB are visible mainly at  $l<500$, combining WMAP data with other CMB datasets (from e.g. ACT or SPT) does not improve this bound. On the contrary, adding  more information on the background cosmological evolution is very useful, because it helps removing degeneracies between the various parameters: in that case, the CMB data can better probe neutrino masses through their background effects. For instance, the bound from WMAP combined with BAO scale measurements and a direct determination of $H_0$ by the Hubble space telescope Key Project \cite{20-Freedman:2000cf}
is significantly stronger: $M_\nu<0.58$~eV~(95\% CL) \cite{20-Komatsu:2010fb}, while the combination of
WMAP with a different $H_0$ determination at various redshift from early-type galaxy evolution gives $M_\nu<0.48$~eV~(95\% CL) \cite{20-Moresco:2012by}.
These are
important results, because they do not depend on the uncertainties from
LSS data discussed next.

\subsection{Large Scale Structure observations}
\label{20-subsec:lss_obs}

The matter power spectrum can be probed with various methods on different scales and redshifts. Let us review here the major techniques which have led so far to relevant neutrino mass bounds.

{\bf Galaxy power spectrum}. Galaxy maps (or, similarly, cluster maps) can be smoothed over small scales and Fourier transformed in order to provide a power spectrum. Relating such a spectrum to the total matter power spectrum is a tricky exercise, especially on small scales (corresponding to wave-numbers $k>0.1\, h\,$Mpc$^{-1}$), because it is difficult to make accurate predictions for non-linear corrections, redshift-space distortions and the light-to-mass bias. Current bounds are based on the analysis of linear scales only, for which the assumption of a scale-independent bias is well motivated. This bias is however left as a free parameter, in such way that galaxy spectrum data give indications on the shape, but not on the global amplitude of $P(k,z)$.
In the next subsection, we will report constraints from the halo power spectrum of Large Red Galaxies (denoted later as Gal-LRG), measured by \cite{20-Reid:2009xm},  the spectrum of the MegaZ catalogue (Gal-MegaZ), used by \cite{20-Thomas:2009ae}, and the WiggleZ Dark Energy Survey (Gal-WiggleZ), used by \cite{20-RiemerSorensen:2011fe}. The first two data sets are actually extracted from the same big survey, the Sloan Digital Sky Survey (SDSS). 

For sufficiently deep galaxy surveys, it is possible to separate galaxies into redshift bins and compute different correlation functions at different redshifts. This technique, called tomography, can be very useful for constraining the scale-dependent growth factor induced by neutrino masses. In that case, the data can be reduced to a set of two-dimension power spectra in different shells, each of them related to $P(k,z)$ in a narrow redshift range. Recently, such a tomographic analysis was used by  \cite{20-Xia:2012na} for constraining neutrinos, using galaxies from the Canada-France-Hawaii-Telescope Legacy Survey (CFHTLS), split in three redshift bins covering the ranges $0.5<z<0.6$, $0.6<z<0.8$ and $0.8<z<1.0$ (this dataset will be denoted as Gal-CFHTLS).

{\bf Cluster mass function.}
Instead of probing directly the matter power spectrum $P(k,z)$ from the spatial distribution of objects, it is possible to constrain integrated quantities of the type $\int dk\, P(k) W(k)$, where $W(k)$ stands for a given window function. One such quantity is related to the histogram of cluster masses. If the mass of a significant number of galaxy clusters within a given redshift bin is known, this histogram gives an estimate of the so-called cluster mass function, $dn(M,z)/dM$, with $dn$ being the number of clusters of redshift approximately equal to $z$, and with a mass in the range $[M, M+dM]$. This function is related to $\sigma^2(M,z)$, the variance of the density in spheres enclosing a mass $M$, itself derived from the convolution of the power spectrum $P(k,z)$ with an appropriate window function. In the next subsection we will refer to bounds derived from cluster abundances probed by X-ray observations from the ROSAT survey, presented by \cite{20-Mantz:2009fw} (denoted later as Clus-ROSAT), and by optical observations from the MaxBCG catalogue, presented by \cite{20-Reid:2009nq} (denoted later as Clus-MaxBCG). 

{\bf Galaxy weak lensing.}
The image of observed galaxies is distorted by gravitational lensing effects, caused by density fluctuations along the line of sight. One of these effects is called cosmic shear.\index{cosmic shear} It corresponds to the squeezing of an image in one direction in the sky, and its stretching in the orthogonal direction. Because such distortions are coherent over the angular size of the lensing potential wells responsible for lensing, they tend to align slightly the apparent major axis of galaxies in a given patch of the sky. Hence the average cosmic shear in a given direction can be estimated statistically by averaging over major axis orientations.
The analysis of a catalogue of images leads to a map of the lensing potential, itself related to the matter power spectrum $P(k,z)$ through the Poisson equation.
If the number of source galaxies is sufficient, it is possible to split the catalogue in several redshift bins, and to obtain a three-dimensional reconstruction of the gravitational potential in our past-line cone, and of $P(k,z)$ at various redshifts. Cosmic shear tomography is particularly useful for measuring neutrino masses, because it can probe the scale-dependent growth factor induced by neutrino masses over an extended range of redshifts.
Current cosmic shear surveys already allow to put bounds on neutrino parameters: we will refer later to results from the CFHTLS presented by \cite{20-Tereno:2008mm} and denoted as WL-CFHTLS. 

{\bf Lyman alpha forests.}
The most luminous and distant compact objects that we can observe are quasars. Some of the photons emitted by quasars interact along the line-of-sight. In particular, a fraction of photons are absorbed at the Lyman alpha wavelength by hydrogen atoms located in the Interstellar Galactic Medium (IGM). The absorbed fraction in a given point of the photon trajectory is proportional to the local density of neutral hydrogen. Because photons are continuously redshifted, absorption in a given point is seen by the observer as a depletion of the spectrum at a given frequency. Hence, inside a limited range called the {\it Lyman alpha forest}, the frequency-dependence of quasar spectra is a tracer of the spatial fluctuations of the hydrogen density along the line of sight. Lyman alpha forests in quasar spectra offer an opportunity to reconstruct the hydrogen density fluctuation along several line-of-sights in a given redshift range. After Fourier expanding each spectrum and averaging over many spectra, one gets an estimate of the  so-called flux power spectrum $P_F(z,k)$, that can be related to the total matter power spectrum $P(k,z)$. Unfortunately, the flux power spectrum does not probe linear scales, but mildly non-linear scales. In order to relate $P_F(z,k)$ to $P(k)$, it is necessary to perform N-body simulations with a hydrodynamical treatment of baryons, accounting for the complicated thermodynamical evolution of the IGM (which depends on star formation). Also, a limitation of this technique comes from the fact that the emitted quasar spectra already have a non-trivial frequency dependence, and that photons are affected by several other effects than Lyman alpha absorption along the line-of-sight. Nevertheless, a careful modeling of all relevant effects allows to obtain interesting constraints. The fact that Lyman alpha forests probe mildly non-linear scales rather than strongly non-linear ones is of course crucial for keeping systematic errors under control. Lyman alpha observations typically constrain the matter power spectrum in the wavenumber range $0.3 < k < 3\,h$/Mpc and in the redshift range $2 < z < 5$.
We mention below some neutrino mass bounds inferred from quasar spectra obtained by the SDSS and presented in \cite{20-Viel:2010bn}, denoted as Ly-$\alpha$-SDSS.

\subsection{Large Scale Structure bounds}
\label{20-subsec:present_GRS}

Using LSS observations in combination with CMB data offers an opportunity to observe (or to bound) the step-like suppression of the matter power spectrum in presence of neutrino masses, as explained in section~\ref{20-subsec:impact_pk} and illustrated in Fig.~\ref{20-fig_tk}. The use of CMB data is crucial in order to constrain parameters like the baryon density, the primordial spectrum amplitude, the tilt, and a combination of $\omega_M$ and $h$. Without such constraints, there would be too much freedom in the matter power spectrum fitted to LSS data for identifying a smooth step-like suppression.

We summarize in Table~\ref{20-chapCMB:tabMASS} the main constraints on $M_\nu$ available at the time of writing, obtained from combinations of CMB plus  homogeneous cosmology data,  galaxy power spectrum data and cluster abundance data. 
Current data set are far from reaching the sensitivity required to probe the mass splitting of the total mass $M_\nu=\sum_i m_{\nu_i}$ between different species. The constraints mentioned below have been derived in the case of three degenerate neutrinos with mass $m_\nu=M_\nu/3$, but they roughly apply to the total mass of any scenario. Also, the bounds shown in the first column of Table~\ref{20-chapCMB:tabMASS} have been obtained under the assumption of a minimal $\Lambda$CDM model with massive neutrinos,  featuring seven free parameters. More conservative bounds are sometimes derived for basic extensions of this model, with one or two more parameters. The constraints do not change significantly when assuming, for instance, a primordial spectrum with a running of the tilt $[d \ln n_s/d \ln k]$, or a significant contribution to the CMB of primordial gravitational waves \cite{20-Reid:2009nq}. Parameters known to be slightly degenerate with neutrino masses and leading to weaker bounds are $w$, the equation-of-state parameter of a Dark Energy component (substituting the cosmological constant), and $N_{\rm eff}$, the effective number of neutrinos discussed in section \ref{20-subsec:neff}. 
The degeneracy with $w$, explained in \cite{20-Hannestad:2005gj}, is illustrated by the last column in Table~\ref{20-chapCMB:tabMASS}. 
\begin{table}[t]
\begin{center}
\begin{tabular}{|l|c|c|c|}
\hline
Cosmological data & Reference & $w=-1$ & $w\neq -1$\\
\hline
WMAP7+Gal-LRG+$H_0$  & \cite{20-Komatsu:2010fb} & 0.44 & 0.76\\
WMAP5+Gal-MegaZ+BAO+SNIa & \cite{20-Thomas:2009ae} & 0.325 & 0.491\\
WMAP5+Gal-MegaZ+BAO+SNIa+$H_0$ & \cite{20-Thomas:2009ae}  &0.281& 0.471\\
WMAP7+Gal-WiggleZ&\cite{20-RiemerSorensen:2011fe}&0.60&--\\
WMAP7+Gal-WiggleZ+BAO+$H_0$&\cite{20-RiemerSorensen:2011fe}&0.29&--\\
WMAP7+Gal-CFHTLS & \cite{20-Xia:2012na} & 0.64(0.44) & --\\
WMAP7+Gal-CFHTLS+$H_0$ & \cite{20-Xia:2012na} & 0.41(0.29)& --\\
\hline
WMAP5+BAO+SNIa+Clus-ROSAT & \cite{20-Mantz:2009fw} & 0.33 & 0.43\\
WMAP5+$H_0$+Clus-MaxBCG & \cite{20-Reid:2009nq} & 0.40 & 0.47\\
\hline
WMAP5+BAO+SNIa+WL-CFHTLS & \cite{20-Tereno:2008mm} & 0.53 & --\\
\hline
\end{tabular}
\end{center}
\caption{95\% CL upper bounds on the total neutrino mass $M_\nu$ in eV, for various combinations of CMB, homogeneous cosmology and LSS data sets. The first seven lines refer to galaxy power spectrum measurements, the next two lines to cluster mass function measurements, the last line to a weak lensing survey. WMAP5 or 7 stands for WMAP 5 or 7-year data. $H_0$ refers to the direct measurement by the HST Key Project \cite{20-Freedman:2000cf}, and BAO to estimates of the scale of Baryon Acoustic Oscillations at various redshifts.  Other acronyms refer to various Large Scale Structure dataset referred in section~\ref{20-subsec:lss_obs}. In the last column, the cosmological constant was replaced by a Dark Energy component with arbitrary equation-of-state parameter $w$.}\label{20-chapCMB:tabMASS}
\end{table}
We did not include in this table current limits from Lyman alpha forest data: this is a delicate matter for the reasons mentioned previously, and a careful modeling of all systematic effects leads to rather weak neutrino mass bounds. The conservative analysis of \cite{20-Viel:2010bn}, based on Ly-$\alpha$-SDSS, gives a bound $M_\nu<0.9$ eV (95\% CL) from Ly-$\alpha$-SDSS data alone. 

In conclusion, the combination of available data sets consistently indicates that  the total neutrino mass is below 0.3 eV at the 95\% CL (0.5~eV if we allow for Dark Energy with arbitrary $w$). This means that the ``degenerate scenario'' in which all neutrinos share roughly the same mass is almost excluded. The data is about to probe the region in which masses are different from each other, and are ranked according to the NH or IH scenario. Other recent summaries of existing bounds have been recently presented in 
\cite{20-Hannestad:2010kz,20-Komatsu:2010fb,20-GonzalezGarcia:2010un,20-Reid:2009nq,20-Abazajian:2011dt}.

\section{Future sensitivities on neutrino masses from cosmology}
\label{20-sec:future}

Future CMB observations will have increasing sensitivity to neutrino masses, not only thanks to smaller error bars on the temperature and polarization spectra. They will make it possible to probe the large scale structure of the universe with a new technique: CMB lensing extraction. The weak lensing of the last scattering surface by nearby galaxy clusters induces specific non-gaussian patterns in CMB maps that can be extracted using some non-linear estimators. This method allows to measure the lensing potential up to $z \sim 3$, and to infer constraints on $P(k,z)$ at such high redshifts.

In the very close future, significant improvements on the neutrino mass bounds will be triggered by the Planck CMB satellite. The forecasts presented in \cite{20-Perotto:2006rj} predict a neutrino mass sensitivity of $\sigma(M_\nu) \sim 0.1$ eV from Planck alone, using the lensing extraction technique of \cite{20-Okamoto:2003zw}. This would be twice better than without lensing extraction. 

Several galaxy surveys with better sensitivity and larger volume are about to release data or have been planned over the next decades, including the Baryon Oscillation Spectroscopic Survey\footnote{\tt cosmology.lbl.gov/BOSS/} (BOSS), the Dark Energy Survey\footnote{\tt www.darkenergysurvey.org/} (DES), the Large Synoptic Survey Telescope\footnote{\tt www.lsst.org/ } (LSST) or the Euclid satellite\footnote{\tt sci.esa.int/euclid}. Also, in \cite{20-Wang:2005vr} it was pointed out that in the future accurate measurements could be inferred from cluster surveys. Because clusters are more luminous than galaxies, they can be mapped up to higher redshift.  Concerning cosmic shear surveys, spectacular improvements are expected from Pan-STARRS\footnote{\tt pan-starrs.ifa.hawaii.edu/ }, or the DES, LSST and the Euclid surveys already mentioned above.

In a near future, the prediction of ref.\ \cite{20-Sekiguchi:2009zs} is that the combination of Planck (with lensing extraction) with BAO scale information from BOSS could lower the error down to $\sigma(M_\nu) \sim 0.06$ eV. In addition, the authors of \cite{20-Gratton:2007tb} find that adding Lyman alpha data from BOSS should lead to comparable sensitivities, and even better results might be expected from the addition of galaxy power spectrum data from the same survey.

With better tomographic data (for either galaxy clustering or cosmic shear), it will become possible to probe the scale-dependence of the growth factor induced by neutrino masses (or in other words, the fact that the step-like suppression has an amplitude increasing with time), and to reach spectacular sensitivites. We present below the typical sensitivities expected for a collection of planned surveys (not all approved). These numbers should be taken with care because forecasts are based on an idealization of each experiment, as well as on several assumptions like the underlying cosmological model, or even the fiducial value of the neutrino mass itself. 

In ref.\ \cite{20-Lahav:2009zr} it was found that the measurement of the galaxy harmonic power spectrum in seven redshift bins by the DES should lead to a sensitivity of $\sigma(M_\nu) \sim 0.06$~eV when combined with Planck data (without lensing extraction). Similar bounds were derived in \cite{20-Namikawa:2010re} for another combination of comparable experiments. This shows that at the horizon of 2014 or 2015, a total neutrino mass close to $M_\nu\simeq0.1$ eV could be marginally detected at the 2-$\sigma$ level by cosmological observations. Because this value coincides with the lowest possible total mass in the inverted hierarchy scenario, the latter could start to be marginally ruled out in case the data still prefers $M_\nu=0$. 

The sensitivity of cosmic shear data from a satellite experiment comparable to Euclid was calculated in \cite{20-Kitching:2008dp}, where it was found that it
would shrink to $\sigma(M_\nu) \sim 0.03$ eV in combination with Planck data (without lensing extraction). The forecast of \cite{20-Carbone:2010ik} based on galaxy clustering data also from Euclid (completed at small redhsift by similar data from BOSS) gives comparable numbers. Constraints based on the ground-based Large Synoptic Survey Telescope should be slightly weaker \cite{20-Hannestad:2006as}. Hence, in the early 2020's, we expect that a combination of cosmological data sets could detect the total neutrino mass of the normal hierachy scenario, $M_\nu\simeq0.05$~eV, at the 2-$\sigma$ level. If the total mass is instead close to $M_\nu\simeq0.1$~eV, it will be detected at the 4-$\sigma$ level. However, in that case, available experiments would not have enough sensitivity for making the difference between an inverted and normal hierarchy scenario with the same $M_\nu$.

Even more progress could be provided by the promising technique of 21-cm surveys.  Instead of mapping  the distribution of hydrogen atoms trough the absorption rate of photons traveling from quasars, it should be possible to observe directly the photons emitted by these atoms at  a wavelength $\lambda\simeq21$ cm from the transition
from one hyperfine level to the other. While travelling towards the observer, these photons are redshifted, and seen with a wavelength indicating the position of the emitting atoms in redshift space.
Recent theoretical progresses in this field show that using this technique, future dedicated experiments should be able to map hydrogen and hence baryonic fluctuations at very high redshift (typically $6<z<12$), and to probe the matter power spectrum deep inside the matter-dominated regime on linear scales \cite{20-Pritchard:2011xb}. This field is still in its infancy, and the forecasts presented so far have to be taken with care, due to the difficulty to make a realistic estimate of systematic errors in future data sets. A
sensitivity of $\sigma(M_\nu) \sim 0.075$~eV for the combination of Planck with the Square Kilometer Array (SKA) project, or $\sigma(M_\nu) \sim 0.0075$~eV with the Fast Fourier Transform Telescope (FFTT), was found in \cite{20-Pritchard:2008wy}. However, the authors show that such impressive experiments would still fail in discriminating between the NH and IH scenario.

An eventual post-Planck CMB satellite or post-Euclid survey would also have a great potential. The forecast analysis in \cite{20-Lesgourgues:2005yv} shows that for 
a CMB satellite of next generation one could get $\sigma(M_\nu) \sim 0.03$ eV alone, thanks to a very precise reconstruction of the CMB lensing potential, while \cite{20-Wang:2005vr} discusses the potential of cluster surveys. Finally, the authors of \cite{20-Jimenez:2010ev} show how far the characteristics of an hypothetical galaxy or cosmic shear survey should be pushed in order to discriminate between two allowed NH and IH scenarios with the same total mass. 

\section{Conclusions}

Neutrinos, despite the weakness of their interactions and their small
masses, can play an important role in Cosmology that we have reviewed
in this contribution. In addition, cosmological data can be used to
constrain neutrino properties, providing information on these elusive
particles that complements the efforts of laboratory experiments. In
particular, the data on cosmological observables have been used to
bound the radiation content of the Universe via the effective number of neutrinos, including a potential extra
contribution from other relativistic particles.

But probably the most important contribution of Cosmology to our
knowledge of neutrino properties is the information it can provide on
the absolute scale of neutrino masses. We have seen that the analysis
of cosmological data can lead to either a bound or a measurement of
the sum of neutrino masses, an important result complementary to
terrestrial experiments such as tritium beta decay and neutrinoless
double beta decay experiments. In the next future, thanks to the data
from new cosmological experiments we could even hope to test the
minimal values of neutrino masses guaranteed by the present evidences
for flavour neutrino oscillations.  For this and many other reasons, we
expect that neutrino cosmology will remain an active research field in
the next years.

\section{Acknowledgments}
SP was supported by the Spanish grants FPA2011-22975 and Multidark Consolider CSD2009-00064
(MINECO), and PROMETEO/2009/091 (Generalitat Valenciana), as well as by the EC contract
UNILHC PITN-GA-2009-237920. He thanks the Galileo Galilei Institute for Theoretical Physics 
for the hospitality and the INFN for partial support during the completion of this work.

\section{References}
\label{20-sec:references}

  \end{document}